\documentstyle[12pt,fleqn,psfig]{article}
\newcommand{\nwc}{\newcommand}
\def\gtrsim{\mathrel{\relax{\raisebox{3pt}{$\mathord{>}$} \kern-0.75em
\raisebox{-2pt}{$\sim$}}}}
\def\lesssim{\mathrel{\relax{\raisebox{3pt}{$\mathord{<}$} \kern-0.75em
\raisebox{-2pt}{$\sim$}}}}

\newcommand{\be}{\begin{equation}}
\newcommand{\ee}{\end{equation}}
\newcommand{\nn}{\nonumber}
\newcommand{\beba}{\begin{equation}\begin{array}{lcl}}
\newcommand{\eaee}{\end{array}\end{equation}}
\newcommand{\bea}{\begin{eqnarray}}
\newcommand{\eea}{\end{eqnarray}}
\newcommand{\ba}{\begin{array}}
\newcommand{\ea}{\end{array}}

\newcommand{\dis}{\displaystyle}
\newcommand{\text}{\textstyle}
\newcommand{\scr}{\scriptstyle}
\newcommand{\sscr}{\scriptscriptstyle}
\newcommand{\ns}{\normalsize}
\newcommand{\refs}[1]{(\ref{#1})}

\nwc{\ra}{\rightarrow}
\nwc{\lra}{\longrightarrow}
\nwc{\lera}{\leftrightarrow}
\nwc{\lolera}{\longleftrightarrow}
\nwc{\pa}{\partial}
\nwc{\pri} {^{\prime}}
\nwc{\dpr} {^{\prime\prime}}
\def\a{\alpha}
\def\b{\beta}
\def\g{\gamma}

\def\d{\delta}

\def\ve{\varepsilon}
\def\f{\phi}
\def\vf{\varphi}

\def\l{\lambda}
\def\m{\mu}
\def\n{\nu}
\def\o{\omega}

\def\th{\theta}

\def\r{\rho}

\def\x{\xi}

\def\F{\Phi}

\def\J{\Psi}

\def\O{\Omega}

\def\Ricci{{\cal R}}

\def\cl{{\cal L}}
\def\cc{{\cal C}}
\def\co{{\cal O}}
\def\bx{{\bf x}}
\def\bk{{\bf k}}
\def\bn{{\bf n}}

\def\zz{\relax{\sf Z\kern-.3em Z}}
\def\ZZ{\relax{\sf Z\kern-.4em Z}}
\def\ZZZ{Z\kern -0.28em Z}
\def\CC{{\rm \kern .25em
             \vrule height1.4ex depth-.12ex width.06em\kern-.31em C}}
\begin{document}
\begin{titlepage}
\title{{\large\bf Scalar field fluctuations in Pre--Big--Bang Cosmologies}\\
                          \vspace{-5cm}
                          \hfill{\ns TUM--HEP 276/97\\}
                          \hfill{\ns SFB--375/179\\}
                          \hfill{\ns MPI--PhT 97--24\\}
                          \hfill{\ns gr--qc/9704083\\[.1cm]}
                          \hfill{\ns April 1997}
                          \vspace{3cm} }

\author{Zygmunt Lalak
        \thanks{Email : lalak@fuw.edu.pl}\\[1cm]
        {\ns Max--Planck--Institut f\"ur Physik}\\
        {\ns F\"ohringer Ring 6}\\
        {\ns D-80805 M\"unchen.}\\
        {\ns Institute of Theoretical Physics}\\
        {\ns University of Warsaw}\\
        {\ns Ho\.za 69, 00--681 Warsaw, Poland\footnote{Permanent address}}
        \and\setcounter{footnote}{6}
        Rudolf Poppe\thanks{Email : rpoppe@physik.tu-muenchen.de}\\[1cm]
        {\ns Physik Department}\\
        {\ns Technische Universit\"at M\"unchen}\\
        {\ns D-85747 Garching, Germany}}
\date{}
\maketitle
\begin{abstract} \baselineskip=6mm\noindent
  Jordan--Brans--Dicke theories with a linearized potential for the scalar
  field are investigated in the framework of the stochastic approach. The
  fluctuations of this field are examined and their backreaction on the
  classical background is described. We compute the mode functions and analyze
  the time evolution of the variance of the stochastic ensemble corresponding
  to the full quantum scalar field in the pre--big--bang regime. We compute
  fluctuations of the term discriminating between the two branches of solutions
  present in the theory. We find, both analytically and upon direct integration
  of the stochastic equations of motion, that the dispersion of these
  fluctuations grows to achieve the magnitude of the term separating the two
  classical solutions. This means that the ensembles representing classical
  solutions which belong to different branches do overlap; this may provide a
  quantum mechanical realization at the level of field theory to change among
  solutions belonging to different branches.
\end{abstract}

\thispagestyle{empty}
\end{titlepage}
\newpage


\setcounter{page}{1}
\section{Introduction}
\label{introduction}

In general scalar tensor theories of gravity, such as Jordan--Brans--Dicke
(JBD) theories, the gravitational coupling constant is replaced by a dynamical
field~\cite{brans}. The theories are parameterized by a dimensionless kinetic
coupling parameter $\o$. In the limit $\o\ra\infty$ these theories go over to
Einstein gravity. From time--delay experiments it is known that $\o$ must be
larger than $\sim 500$ at the present epoch~\cite{reasenberg}. It is
conceivable that in the early universe the actual value of $\o$ differed
significantly from that bound in being itself a function of the Planck mass,
i.~e. of the JBD field $\F = M_{p}^{2} = G^{-1}$ (see e.~g. ~\cite{levin}).
Being part of the dynamical evolution in the early stages $\F$ is assumed to
finally acquire a finite value which is consistent with observations of $G$
and $\o$ today. \\ 
Theories of gravity arising from Kaluza--Klein theories or supergravity may
appear as a JBD theory after compactification to four dimensions with $\o$ of
the order of unity~\cite{freund} with the square root of the metric of the
compactified dimensions acting as a JBD field on the uncompactified
space--time manifold. On the other hand, the universal part of the low energy
effective string action~\cite{eff_action} can be viewed as a JBD theory with
the specific choice $\o =-1$ after a field redefinition $\F =e^{-\f}$ and $\f$
being the dilaton of string theory. In this context a (nonperturbatively)
generated potential for the corresponding scalar field is taken into account
which can effectively act as a cosmological constant and drive a period of
inflation. This has been extensively studied in the
literature~\cite{inflation}. Recently the possibility of having a kinetic type
of inflation (in contrast to that driven by a potential term of the JBD field
or an inflaton) has been an active field of investigation~\cite{levin,la}. \\ 
Especially much work has been devoted to the special case of the string
low energy effective action ($\o =-1$). It has been observed that there are
two different cosmological solutions (one defined for $t>t_{0}$ and the other
for $t<t_{0}$), related by scale factor duality (SFD) and separated from each
other by a curvature singularity, which offer a quite attractive
pre--big--bang scenario~\cite{ven}. Due to the negative kinetic coupling one
of the pre--big--bang solutions undergoes a superinflationary period totally
driven by the kinetic energy term. This stage is supposed to be followed after
the big bang by an ordinary radiation dominated expanding Robertson--Walker
universe. \\
On tree--level, the curvature singularity prevents a smooth transition from
the inflationary phase to the radiation dominated universe (graceful exit
problem)~\cite{bru_ven}. Later it was proven~\cite{kal_mad} that a dilaton
potential cannot trigger such an exit. However, the singularity is thought to
be a property of the low energy effective action which would be absent if
higher order corrections were taken into account thus intertwining these two
phases~\cite{east_maed}. As an alternative approach quantum cosmological
techniques have been applied to the scenario in the regime where the
tree--level effective action is still considered to be a good
approximation~\cite{quantcos}. \\ 
SFD is part of a larger $O(d,d)$--symmetry~\cite{odd} present in low energy
action and reduces to SFD in the case of a homogeneous and isotropic universe.
A potential for the dilaton as well as a curvature term break SFD. In many
situations one can have an approximate symmetry since a spatial curvature term
can safely be ignored at very early times and the influence of the dilaton
potential is suppressed by the kinetic term as in the case of the
superinflationary branch close to the curvature singularity. General scalar
tensor theories such as the JBD theory don't possess this kind of
$O(d,d)$--symmetry which is clearly of stringy origin. Nevertheless solutions
can be found which form two disconnected branches separated by a field
dependent discriminant; they are (apart from singularities) defined for all
times such that one always finds similar pre--big--bang scenarios as those
described above for the stringy case, but one also faces similar problems. In
JBD theories, i.~e. those with arbitrary but constant $\o$, such periods of
kinetic inflation are present only if the parameter $\o$ is
negative~\cite{levin}. These models will be the most interesting ones for our
purpose~\cite{lidsey}. \\ 

Quantum fluctuations in JBD theory have been studied on inflationary
backgrounds, i.~e. those fulfilling the slow--roll conditions,
in~\cite{garcia}, where also density perturbations have been computed; similar
analyses have been performed on more general Robertson--Walker backgrounds,
but without potential, or alternatively for an exponential potential in the
context of extended inflation in the Einstein frame~\cite{mallik}. For the
string cosmological scenario ($\o=-1$) scalar and tensor perturbations have
been computed~\cite{gravrad} and the exit problem via quantum backreaction has
been investigated in~\cite{antoniadis}. We point out that a full analysis of
the backreaction problem in JBD theory on general cosmological backgrounds
including a generic potential (and also a matter contribution to the energy
density) is still lacking. One step into that direction shall be the subject
of this paper. For the problem to be tractable analytically one could e.~g.
expand the solutions for small values of the potential or for small times
where the influence of the potential can be linearized. \\ 

The intention of this paper is to study the dynamics of quantum fluctuations
of the JBD field and its backreaction on the background fields by means of a
stochastic analysis. For this purpose we split the field and its first time
derivative (which will be treated as an independent stochastic variable) into
sub-- and super--horizon sized modes. This technique has been applied to
similar problems before~\cite{rey}. In this framework it is possible to trace
the time evolution of the long wavelength (coarse grained) fields which we
interpret as the classical background under the influence of the stochastic
noise operators. It is possible to write down a system of stochastic
differential equations (Langevin equations) and the corresponding
Fokker--Planck equation for the probability distribution of the stochastic
variables, i.~e. the classical background field and its first time derivative.
The Langevin equations can be integrated numerically by a well known
method~\cite{risken,yi}. For the Fokker--Planck equation one can in principle
search for an analytic solution. \\ 
The stochastic method is suitable to investigate the effects of quantum
effects on the background quantities as long as the size of the fluctuations
are under control. It is a well defined technique and can serve as an
alternative to the minisuperspace approach of examining the exit problem of
string cosmologies. Since the stochastic differential equations can be
integrated for any form of the potential one can hope to make statements about
the influence of the scalar field potential in this context. However, we
argue that a linearized potential (which shows up as a cosmological term in
the action) will be sufficient as a first approximation. For the linear
potential we will also investigate the mode functions on that background and
their dependence on the potential and the parameter $\o$ since they will
finally enter the stochastic differential equations. By comparing the relative
size of the mode functions and the zero mode it is also possible to check the
validity of the procedure. \\
With the above tools we are able to study the backreaction of the fluctuations
on the dynamical level. We check up on the r\^ole of a large potential and
examine how it decouples from the dynamics in the limit when it becomes small.
We compute the variance of the scalar field fluctuations and its
backreaction on the Hubble parameter which we will finally compare with its
discriminant separating the two different background solutions. If both are
approximately equal, it is no longer possible to unambiguously talk about two
distinct branches. \\

In section~\ref{basics} we give the basic equations of JBD cosmology and fix
our conventions. The setup of the stochastic treatment of the scalar field
fluctuations are presented in section~\ref{stochasticeoms}. We discuss the
general background solutions for a linear potential in JBD cosmology and give
expressions for the mode functions on that background in section~\ref{linpot}
together with numerical solutions for variable $\o$ and strengths of the
potential; we analyze the exit problem of the superinflationary branch in the
presence of quantum fluctuations of a scalar field. The results are summarized
in section~\ref{conclusion}.


\section{Basics of Jordan--Brans--Dicke cosmology}
\label{basics}

We start with the following action of the JBD theory,
\be
S=\frac{1}{16 \pi }\int d^{4}x\sqrt{-g}\left[\F\,\Ricci - 
\o\,\frac{\pa_{\m}\F\,\pa^{\m}\F}{\F} - V(\F) + 16\pi\cl_{m}
\right] \ ,
\label{action} 
\ee
including matter and a finite potential energy term for the JBD scalar field
$\F$. $\Ricci$ is the Ricci scalar of the metric. In the following we keep
$\o$ as an arbitrary parameter of the theory assuming that it is independent
of the field $\F$. Since the physics is more obvious in the JBD frame, we 
will not change into the conformally related Einstein frame. \\ 
Varying the action~\refs{action} with respect to $g^{\mu\nu}$ and $\F$ yields
the following two equations of motion
\bea
R_{\mu\nu} -\frac{1}{2}\,\Ricci\,g_{\mu\nu} & = &  
\frac{8\pi}{\F}\, T_{m\,\mu\nu} + \frac{\o}{\F^{2}}(\pa_{\mu}\F)(\pa_{\nu}\F)
+ \frac{1}{\F}\,\nabla_{\mu}\pa_{\nu}\F \nn \\
& - & g_{\mu\nu}\,\left[\frac{\Box\F}{\F} + \frac{\o}{2\F^{2}}
(\pa_{\a}\F)(\pa^{\a}\F) + \frac{V(\F)}{2\F}\right] \ , \label{eom1} \\ \nn \\
\Ricci & = &  \o\,\frac{(\pa_{\mu}\F)(\pa^{\mu}\F)}{\F^{2}}
- 2\o\frac{\Box\F}{\F} + \frac{\pa\, V(\F)}{\pa\F} \ ,
\label{eom2}
\eea 
where $T_{m\,\mu\nu}$ is the energy--momentum tensor of the matter sector.
Combining the trace of equation~\refs{eom1} with equation~\refs{eom2} one
arrives at a Klein--Gordon equation for the JBD scalar $\F$,
\be
\Box\F = \frac{8\pi}{2\o +3}\, T_{m}
       - \frac{2}{2\o +3}\, V(\F)
       + \frac{1}{2\o +3}\,\frac{\pa V(\F)}{\pa\F}\,\F \ ,
\label{KleinGordon}
\ee
with $T_{m}\equiv T^{\,\,\m}_{m\,\,\m}$ being the trace of the energy--momentum
tensor. For the homogeneous and isotropic metric we use an ansatz of the
Robertson--Walker type defined by
\be
g_{\m\n}\, dx^{\m} dx^{\n} = dt^2 - a^2 (t)\, d\O^{2}_{3} (k) \ .
\label{metric}
\ee 
Here $a(t)$ is the scale factor of the universe and $d\O^{2}_{3} (k) $ stands
for the volume element of the 3--dimensional spacelike hypersurface of
constant $t$ and $k=\pm 1, 0$. In this metric an ideal fluid is described in
its rest frame by $T_{m}=\r - 3p$ (with $\r$ and $p$ denoting the matter
energy density and pressure respectively). The equations~\refs{eom1}
and~\refs{eom2} in the metric~\refs{metric} consist of only two linearly
independent equations, the third one serving as a constraint. We choose the
0--0 component of~\refs{eom1} in the metric~\refs{metric} which reads
\be
H^{2} = \frac{8\pi\r}{3\F}
      + \frac{\o}{6}\,\frac{\dot{\F}^{2}}{\F^{2}} 
      - H\,\frac{\dot{\F}}{\F}
      - \frac{V(\F)}{6\,\F} 
      + \frac{\o}{6\, a^{2}}\frac{(\nabla\F )^{2}}{\F^{2}}
      + \frac{1}{3\, a^{2}}\frac{\triangle\F }{\F}
      - \frac{k}{a^{2}} \ ,
\label{Einstein}
\ee
where we have used the definition of the Hubble parameter, $H\equiv\dot{a} /a$
and an overdot denoting a time derivative. For later purpose it will be
useful to write~\refs{Einstein} as an explicit equation for the Hubble
parameter,
\be
H = -\frac{1}{2}\,\frac{\dot{\F}}{\F}\,\pm\sqrt{
     \frac{8\pi\r}{3\F} 
     + \frac{2\o +3}{12}\,\frac{\dot{\F}^{2}}{\F^{2}} 
     - \frac{V(\F)}{6\,\F}
      + \frac{\o}{6\, a^{2}}\frac{(\nabla\F )^{2}}{\F^{2}}
      + \frac{1}{3\, a^{2}}\frac{\triangle\F }{\F} 
      - \frac{k}{a^{2}}} \ ,
\label{twobranches}
\ee
which shows that there are two different background solutions separated by a
discriminant given by the square root in~\refs{twobranches}. By setting
$\o=-1$ for a homogeneous and free field in a universe described by $\r =k=0$
it becomes apparent that the $+$ sign ($-$ sign) in~\refs{twobranches}
corresponds to the $(-)$ branch ($(+)$ branch) of reference~\cite{bru_ven}
defined only for $t>0$ ($t<0$). Therefore, the discriminant distinguishes
between a superinflationary and a collapsing phase of the $(+)$ branch for
negative times and between an ordinary radiation dominated expanding universe
and a collapsing one of the $(-)$ branch for positive times. At each instant
of time the separation originating from the square root of ~\refs{twobranches}
is, however, equal to the separation between the two different branches if
they {\em were} defined for all times. In subsection~\ref{dispersion} we
shall compare the discriminant with the fluctuations in the JBD field.

Equivalently, the Klein--Gordon equation~\refs{KleinGordon} in the
Robertson--Walker metric~\refs{metric} is given by
\be
\ddot{\F} + 3H\dot{\F} - \frac{\triangle\F}{a^{2}} 
            - \frac{8\pi}{2\o +3}\, T_{m}
            + \frac{2}{2\o +3}\, V(\F)
            - \frac{1}{2\o +3}\,\frac{\pa V(\F)}{\pa\F}\,\F = 0 \ .
\label{KG}
\ee
The system of equations~\refs{Einstein}--\refs{KG} contains also spatial
inhomogeneities of the scalar field $\F(\bx ,t)$ which we assume to be small
but non negligible and living in a homogeneous universe described by the scale
factor $a(t)$. Since the purpose of this paper is to study the behaviour of
inhomogeneous modes of $\F$ we separate them from the homogeneous background,
i.~e. the zero mode of $\F$. We decompose the scalar field $\F$ into a
homogeneous background and an inhomogeneous part according to
\be \F ({\bf x},t) = \F_{0} (t) + \d({\bf x},t) \ ,
\label{split1}
\ee
and expand the latter within a box of volume $v=l^3$ into
its modes as follows~:
\be
\d({\bf x},t) = l^{-3/2}\,\sum_\bk\left[ a_{\bk}\,\vf_{\bk} (t)\, 
                  e^{i\bk \cdot \bx}
                + a_{\bk}^{\dagger}\,\vf_{\bk}^{\ast} (t)\, 
                  e^{-i\bk \cdot \bx}\right] \ .
\label{fourier}
\ee  
The sum is performed over all wave vectors $\bk = \frac{2\pi\bn}{l}\ne 0$ with
$\bn\in\ZZ^3$. The equation of motion for the homogeneous zero mode $\F_{0}
(t)$ is clearly given by~\refs{KG} with the spatial derivative dropped.
On the other hand, substituting the ansatz~\refs{split1} into the Klein--Gordon
equation~\refs{KG}, expanding to linear order in $\d({\bf x},t)$ and using the
equations of motion for the background $\F_{0} (t)$ we immediately arrive at
the equation of motion for the mode functions $\vf_{\bk} (t)$,
\be
\ddot{\vf}_{\bk} \, + \, 3H\,\dot{\vf}_{\bk}  
   \, + \,\left[ \frac{|\bk |^{2}}{a^{2}} + \frac{1}{2\o +3}\,\bigg(
   V^{\prime}(\F_{0}) - V^{\prime\prime}(\F_{0})\,\F_{0} \bigg)
\right] \vf_{\bk}  = 0 \ ,
\label{eom_modes}
\ee 
with $V^{\prime}(\F_{0})$ denoting $\pa V(\F)/\pa\F |_{\F=\F_{0}}$, etc.
This equation is independent of the curvature term $k/a^{2}$ which
means that the topology of the spatial section enters only via the
background quantities. \\

For the rest of the paper we will neglect the spatial curvature term ($k=0$)
as well as the contribution of ordinary matter to the energy density ($\r
=p=0$). For completeness we give the background solutions in the case where in
addition the potential is neglected. The equations of motion~\refs{Einstein}
and~\refs{KG} allow the following power--law solutions which are defined for
all times except $t=0$:\footnote{the singularity can certainly also be at
  $t=t_{0} $. For simplicity we set $t_{0}=0 $.}
\be
\F_{0}\, (t) \sim |\, t\, |^{r(\o)}
\qquad\mbox{and}\qquad 
a (t) \sim |\, t\, |^{q(\o)} \ ,
\label{V0solutions}
\ee
with the exponents given by
\be
r(\o) = \frac{1\pm\sqrt{9+6\o}}{4+3\o}
\qquad\mbox{and}\qquad 
q(\o) = \frac{1-r(\o)}{3} \ .
\label{exponents1}
\ee
After the field redefinition $\F =e^{-\f}$ we recover the known result for the
low energy string action ($\o =-1$) \cite{bru_ven}
\be
\f (t)\sim (\mp \sqrt{3} -1)\,\ln |\, t\, |
\qquad\mbox{and}\qquad 
a(t)\sim |\, t\, |^{{\text \mp \frac{1}{\sqrt{3}}}} \ , 
\label{ven1}
\ee
in which case it is invariant under SFD. Comparing \refs{exponents1} with
\refs{ven1} it is clear that the $+$ sign in front of the square root in
\refs{exponents1} corresponds to the accelerated expansion in pre--big--bang
cosmology, i.~e. the superinflationary part of the $(+)$ branch. For $t>0$ the
part of the $(-)$ branch which corresponds to the $-$ sign in \refs{exponents1}
can be smoothly connected to a FRW universe with decelerated expansion and can
therefore account for our present universe. We note that there is also a
constant background solution for the vanishing potential, i.~e. $\F_{0} =H=0$.
We will not be interested in that possibility, however, since it implies a
static universe.

For the zero potential background the mode functions can be directly read off
from equation~\refs{eom_modes} and given in terms of Hankel functions,
\be
\vf_{\bk} (t) \propto |\, t\, |^{-\a(\o)}\, \cc_{\n(\o)}  
\left( \b(\o)\, |\, t\, |^{\g(\o)} \right) \ ,
\label{modes1}
\ee
where for shorthand we introduced the quantities
\be
\!\a(\o) = -\frac{r(\o)}{2}, \quad
\!\b(\o) = \frac{1}{1\! -\! q(\o)}\frac{|\bk |}{a_{0}}, \quad
\!\g(\o) = 1\! -\! q(\o), \quad
\!\n(\o) = \pm\frac{1}{2}\frac{1\! - \! 3q(\o)}{1\! -\! q(\o)}  \ .
\label{index}
\ee 

By $a_{0}$ we denote any initial value for the scale factor which we can set
to one for simplicity. $\cc_\n $ stands for any linear combination of the
Hankel functions $H_\n^{(1)}$ and $H_\n^{(2)}$ with the two integration
constants $c_{1}(\bk )$ and $c_{2}(\bk )$ multiplying the respective Hankel
functions. Choosing a certain linear combination corresponds to making a
specific choice for the vacuum state; it is subject to the Klein--Gordon
normalization condition $|c_{2}(\bk )|^{2} \, - |c_{1}(\bk )|^{2} = 1$. The
correct Minkowski limit (positive frequencies) can be obtained by the simplest
choice $c_{1}=0$ and $c_{2}=1$ for all wavenumbers which is the vacuum state
we shall adopt in the following. Choosing the negative sign in the above
expression for $\n(\o)$ gives the non trivial solution $\F_{0}\sim |\, t\,
|^{r(\o)}$ in the limit $\bk\rightarrow 0$, whereas the positive sign yields
the constant solution.


\section{Stochastic equations and correlation functions}
\label{stochasticeoms}
In this section we will derive the modified equations of motion in the
stochastic approach and the correlation functions between the various
stochastic field operators. For this purpose we split the field and its first
time derivative in sub-- and super--horizon sized parts~\cite{rey,yi},
\begin{equation}\begin{array}{lllll}
\F({\bf x}, t) & = & \F_{<} ({\bf x}, t) & + & \F_{>} ({\bf x}, t) 
\vspace{2mm}\\
\dot{\F}({\bf x}, t) & = & v_{<} ({\bf x}, t) & + & v_{>} ({\bf x}, t) \ ,
\label{split2}
\end{array}\end{equation}
and decompose each of them into Fourier modes according to
\bea
\F_{> \atop <} ({\bf x}, t) & = & l^{-3/2}\,\sum_\bk \th [\pm\, (k-\ve aH)]
\left( a_{\bk}\,\vf_{\bk} (t)\, e^{i\bk \cdot \bx}
+ a_{\bk}^{\dagger}\,\vf_{\bk}^{\ast} (t)\, e^{-i\bk \cdot \bx}\right) 
\label{splitting1}\vspace{2mm}\\
v_{> \atop <} ({\bf x}, t) & = & 
l^{-3/2}\,\sum_\bk \th [\pm\, (k-\ve aH)]
\left( a_{\bk}\,\dot{\vf}_{\bk} (t)\, e^{i\bk \cdot \bx}
+ a_{\bk}^{\dagger}\,\dot{\vf}_{\bk}^{\ast} (t)\, e^{-i\bk \cdot 
\bx}\right) \ .
\label{splitting2}
\eea
The quantities $\F_{<}$ and $v_{<}$ are the long--wavelength (coarse--grained)
fields and correspond to the {\em lower} sign in the $\th$ functions of
equation~\refs{splitting1} and~\refs{splitting2}, whereas $\F_{>}$ and $v_{>}$
are the short--wavelength fluctuations (referring to the {\em upper} sign).
The parameter $\ve$ denotes the splitting point and should be of the order of
unity if we choose the horizon length as our effective physical cutoff.
We define the stochastic
noise operators
\bea
\chi\, ({\bf x}, t) & = & 
\ve a(H^{2}+\dot{H})\, l^{-3/2}\,\sum_\bk \d (k-\ve aH)
\left( a_{\bk}\,\vf_{\bk} (t)\, e^{i\bk \cdot \bx}
+ a_{\bk}^{\dagger}\,\vf_{\bk}^{\ast} (t)\, e^{-i\bk \cdot \bx}\right) 
\nn \\
\x\, ({\bf x}, t) & = & \ve a(H^{2}+\dot{H})\,
l^{-3/2}\,\sum_\bk \d (k-\ve aH)
\left( a_{\bk}\,\dot{\vf}_{\bk} (t)\, e^{i\bk \cdot \bx}
+ a_{\bk}^{\dagger}\,\dot{\vf}_{\bk}^{\ast} (t)\, e^{-i\bk \cdot 
\bx}\right) \ .
\label{noiseop}
\eea
In the following we will aim at deriving a set of stochastic
differential equations (Langevin equations) describing the system given 
by the action \refs{action}.
We regard the short--wavelength quantities $\F_{>} ({\bf x}, t)$
and $v_{>} ({\bf x}, t)$ as small perturbations around the coarse--grained 
fields $\F_{<} ({\bf x}, t)$ and $v_{<} ({\bf x}, t)$ which
we will finally interpret as semiclassical random variables whose
evolution we follow under the influence of the stochastic noise
terms \refs{noiseop}. It is easy to see that
\be
\dot{\F}_{<} ({\bf x}, t) = v_{<} ({\bf x}, t) + \chi\, ({\bf x}, t) \ ,
\label{stoeom1}
\ee
where we have used the relation
\be
\frac{\pa}{\pa t} \,\th [\,\mp (k-\ve aH)] = 
\pm\,\ve a(H^{2} +\dot{H} ) \,\d (k-\ve aH) \ .
\label{theta}
\ee
In the same way we get $\dot{\F}_{>} ({\bf x}, t) = v_{>} ({\bf x}, t) -
\chi\, ({\bf x}, t)$ and hence $\dot{\F} ({\bf x}, t) = v_{>} ({\bf x}, t) +
v_{<} ({\bf x}, t) $. Equation~\refs{stoeom1} is the first stochastic
differential equation since it contains only coarse--grained fields (our
random variables) and the stochastic noise. \\ 
It is more difficult to obtain a Langevin equation for $v_{<} ({\bf x}, t)$. We
write $\dot{v}_{<} ({\bf x}, t) = \ddot{\F} ({\bf x}, t)-\dot{v}_{>} ({\bf x},
t)$ and get
\bea
\dot{v}_{<} ({\bf x}, t) &=&
 -3H\dot{\F} ({\bf x}, t) + \frac{\triangle\,\F ({\bf x}, t)}{a^{2}}
 -  \frac{1}{2\o +3}\, \bigg(\, 2V(\F ) -V\pri (\F ) \,\F ({\bf x}, t)\,\bigg)
\label{vdot1}\nn \\ \\
& & + \,\x\, ({\bf x}, t) - l^{-3/2}\,\sum_\bk\,\th (k-\ve aH)\,
                \left[ a_{\bk} \,\ddot{\vf}_{\bk} (t)\, e^{i\bk \cdot \bx}
                + a_{\bk}^{\dagger} \,\ddot{\vf}_{\bk}^{\ast} (t)\, 
                e^{-i\bk \cdot \bx}\right] \ , \nn
\eea
where we used the equation of motion for the full (inhomogeneous) quantum
field~\refs{KG} and the definition of $v_{>} ({\bf x}, t) $,
equation~\refs{splitting2}. The first line of equation~\refs{vdot1} can be
split into the coarse--grained and short wavelength fields according
to~\refs{split2}. We will expand the potential and its derivative around the
coarse--grained fields,
\be
2V(\F) - V\pri (\F)\,\F =
2V_{0} - V\pri_{0}\,\F_{<} + \Big[\,V\pri_{0} -V\dpr_{0}\,\F_{<} 
\,\Big]\,\F_{>} + \co (\F_{>}^{2}) \ , 
\label{expand}
\ee
where we have used the notation $V_{0}\equiv V(\F_{0})$,
$V_{0}^{\prime}\equiv\pa V(\F)/\pa\F |_{\F=\F_{0}}$, etc. To simplify the
expressions we will approximate $\F_{0}\approx\F_{<}$ and $v_{0}\approx v_{<}$
which means that effectively we neglect all spatial gradients of the
coarse--grained field $\F_{<}$. Later we will see that this is a very good
approximation; it gets exact in the limit $\ve\rightarrow 0$. This
identification will cancel the terms linear in the small quantities $\F_{>}$
and $v_{>}$ in equation~\refs{vdot1}. In the second line of~\refs{vdot1} the
second time derivatives of the mode functions are eliminated via their
equation of motion~\refs{eom_modes}. \\ 
This together with the approximation mentioned above gives the following
differential equation for the coarse--grained field $v_{<}$,
\be 
\dot{v}_{<} ({\bf x}, t) = -3Hv_{<} ({\bf x}, t) - \frac{1}{2\o +3}\,
\bigg(\, 2V_{0} -V\pri_{0} \,\F_{<} ({\bf x}, t)\,\bigg) + \x\, ({\bf x}, t)
\label{stoeom2}
\ee
up to second order in the small quantities. \\
The equations~\refs{stoeom1} and~\refs{stoeom2} constitute a system of
stochastic differential equations for the fields $\F_{<} ({\bf x}, t)$ and
$v_{<} ({\bf x}, t)$ which we regard as independent random variables subject
to the two independent stochastic forces $\chi ({\bf x}, t)$ and $\xi ({\bf
  x}, t)$. Effects of spatial variability of the fields enter only through the
stochastic noise operators. For a full description of the system an equation
similar to~\refs{twobranches} should be added to the two Langevin equations;
the total energy density we assume to be stored mainly in the background
fields. Therefore, we neglect energy contributions arising from the small
quantities $\F_{>} ({\bf x}, t)$ and $v_{>} ({\bf x}, t)$,
\be
H = -\frac{1}{2}\,\frac{v_{<}}{\F_{<}}\,\pm\sqrt{
       \frac{2\o +3}{12}\,\frac{v_{<}^{2}}{\F_{<}^{2}} 
     - \frac{V_{0}}{6\,\F_{<}}} \ .
\label{stochHubble}
\ee
The set of equations~\refs{stoeom1}, \refs{stoeom2} and~\refs{stochHubble} can
be integrated by modeling the stochastic noise terms by a random number
generator. Statistical independence of the numbers requires a Gaussian
distribution. Integrating Langevin equations of that type is a well known
problem and one can use standard techniques~\cite{risken}.

The correlation functions between the stochastic noise operators can be
calculated in a straightforward way,
\bea
\langle\, 0\, |\,\chi\, ({\bf x}, t)\,\chi\, ({\bf x\pri }, t\pri )\,
|\, 0\,\rangle & = &
\frac{1}{2\pi^{2}}\,\ve^{3} a^{3} H^{4} \left( 1+\frac{\dot{H}}{H^{2}}
\right)\,\frac{\sin (\ve aH|\bf{x} -\bf{x}\pri |)}
{\ve aH|\bf{x} -\bf{x}\pri |} \,\,\times \label{correlator} \nn \\ \\
 & & | \vf_{\bk} (t)|^{2} \,\bigg|_{|\bk |=\ve a|H|} \,\d (t-t\pri) \nn \ ,
\eea
and similar expressions for the correlation functions
$\langle\, 0\, |\,\x\, ({\bf x}, t)\,\x\, ({\bf x\pri }, t\pri )\,
|\, 0\,\rangle$ and the cross correlation functions 
$\langle\, 0\, |\,\chi\, ({\bf x}, t)\,\x\, ({\bf x\pri }, 
t\pri )\, |\, 0\,\rangle$ and $\langle\, 0\, |\,\x\, ({\bf x}, t)\,\chi\, 
({\bf x\pri }, t\pri )\, |\, 0\,\rangle$, for which $|\vf_{\bk} (t)|^{2} $
is replaced by $|\dot{\vf}_{\bk} (t)|^{2}$,
$\vf_{\bk} (t)\,\dot{\vf}_{\bk}^{\ast} (t)$ and
$\dot{\vf}_{\bk} (t)\,\vf_{\bk}^{\ast} (t)$ respectively.
In the same way one derives the only non trivial commutator for the noise 
operators,
\bea
[\,\chi\, ({\bf x}, t), \,\x\, ({\bf x\pri }, t\pri )\, ] & = & 
\frac{1}{2\pi^{2}}\,\ve^{3} a^{3} H^{4} \left( 1+\frac{\dot{H}}{H^{2}}
\right)\,\frac{\sin (\ve aH|\bf{x} -\bf{x}\pri |)}
{\ve aH|\bf{x} -\bf{x}\pri |} \,\,\times \label{commutator} \nn \\ \\
 & & \left(\vf_{\bk} (t)\,\dot{\vf}_{\bk}^{\ast} (t) - \dot{\vf}_{\bk}
   (t)\,\vf_{\bk}^{\ast} (t) \right) \,\bigg|_{|\bk |=\ve a|H|} \,
   \d (t-t\pri) \nn \ .
\eea


\section{Mode functions with a cosmological term}
\label{linpot}
The aim of the following section is to analyze the mode functions of the JBD
field on a background evolving under a linearized version of the potential $V(\F)$ in the
action~\refs{action}.
This linear term can be interpreted as a non vanishing cosmological constant,
or as a mass term for the canonical mass dimension 1 field $\chi$ defined 
through $\chi^2 = \F$.
We choose a linear potential first of all because it is
tractable analytically. Moreover, since our main point of interest is the
superinflationary branch before the singularity, it is not the minima of the potential, 
which are modeled by nonlinear terms, which control inflation. Actually,
the influence of the
potential on the dynamics is naively expected to be subdominant, because, for small
values of $\o$,  the grander
part of the energy is stored in the kinetic term of the scalar field. However, the 
strength of the potential can affect the rate of evolution which in turn can change the 
magnitude of quantum (represented here as stochastic) fluctuations. To study this 
effect in the system  not too far away from the singularity the 
linearized version of the potential term 
is expected to serve  sufficiently well. 
The mode functions enter e.~g. in the correlation
functions~\refs{correlator} and the equations for the variance of the
fluctuations. Background solutions for a finite cosmological constant have
been given for the first time in~\cite{uehara} and applied to pre--big--bang
cosmologies in~\cite{lidsey}. For completeness we give a short review of the
derivation in the following subsection.

\subsection{Background solutions for non zero cosmological constant}
\label{backlinpot}
Using the ansatz $V(\F)=-2\l\F$ for the linearized potential the
0--0~component and the i--i~component of the Einstein equation~\refs{eom1} and
the Klein--Gordon equation~\refs{KleinGordon} read respectively (spatial
curvature and matter contributions are omitted)
\bea
3H^{2} - \l & = & \frac{\o}{2}\,\J^{2} - 3H\J \label{coseom1} \\
-2\frac{\ddot{a}}{a} - H^{2} + \l & = & \frac{\o}{2}\,\J^{2} +
\frac{\ddot{\F}_{0}}{\F_{0}} + 2H\J \label{coseom2} \\
\frac{\ddot{\F}_{0}}{\F_{0}} + 3H\J & = & \frac{2\l}{2\o +3} \label{coseom3} \ , 
\eea
where we have used the definitions $H=\dot{a}/a$ and $\J=\dot{\F}_{0}/\F_{0}$.
Defining the following functions
\be
f=\F_{0}\, a^{3}\qquad h=\dot{\F}_{0}\, a^{3}\qquad g=\ddot{\F}_{0}\, a^{3}
\label{functions}
\ee 
and noting that $\ddot{a}/a = \dot{H} + H^{2}$ and $\ddot{\F}_{0}/\F_{0} = \dot{\J} +
\J^{2}$, we find that the linear combination of equations~\refs{coseom1} --
\refs{coseom2} -- $\frac{1}{3}$~\refs{coseom3} reduces to the familiar form
\be 
\ddot{f} - b^{2} f =0 \qquad\mbox{where}\qquad b^{2}=2\l\,\frac{3\o +4}{2\o
  +3} \ .
\label{oscequ}
\ee
The general solution of this equation is $f = A_{+}\; e^{b\, t} + A_{-}\;
e^{-b\, t}$ with constants of integration $A_{+}$ and $A_{-}$ to be determined
later. The lhs of equation~\refs{coseom3} is equal to $\dot{h}/f$ so that it
can be written as
\be
\dot{h} - d^{2}\, f =0 \qquad\mbox{where}\qquad d^{2}=\frac{2\l}{2\o +3} \ ,
\label{equ2}
\ee
with the first integral $h=C + d^{2}\,\int fdt$ ($C$ being an arbitrary
constant). From~\refs{functions} we see that $\dot{f} = h + 3Hf$ which can be
solved for H and since the functions $f$ and $h$ are already known it can be
integrated to yield the scale factor $a$, 
\be
H=\frac{\dot{f} - h}{3f} \qquad\mbox{and}\qquad
a=\tilde{a}_{0}\, |f|^{\frac{1}{3}}\,
\exp\left[-\frac{1}{3}\,\int\frac{h}{f}\, dt\right] \ .
\label{equ3}
\ee
From~\refs{functions} we can read off that $\J = \dot{\F}_{0}/\F_{0} = h/f$ and by
integration we get 
\be
\F_{0} = \tilde{\F}^{\sscr (0)}_{0}\,\exp\left[\int\frac{h}{f}\, dt\right] \ . 
\label{equ4}
\ee

In general the Hubble parameter $H$ will have a singularity. Without loss of
generality we can fix it to be at $t=0$; this requires to choose $A_{+}= -
A_{-}\equiv\pm A$ with $A>0$. Then $f = \pm 2A\sinh (b\, t)$ and the sign
ambiguity determines the two branches of solutions. It turns out that this
fixes the ratio $C/A$ as well meaning that the solutions for $H$, $a$ and $\F_{0}$
that we get by successive integration of the above equations are only
fulfilled if $C/A$ takes the value 
\be
\frac{C}{A} = 2\sqrt{\frac{6\l }{3\o +4}} \ .
\label{C/A}
\ee
We fix the constants of integration $\tilde{a}_{0}$ and $\tilde{\F}^{\sscr (0)}_{0}$ by
making contact with the well known results for $\o =-1$ in the limit
$\l\rightarrow 0$, $a=a_{0}|\, t\, |^{\pm 1/\sqrt{3}}$ and $\f_{0} =\f^{\sscr (0)}_{0}
-(1\mp\sqrt{3})\,\ln|\, t\, |$, $\f_{0}$ being again the stringy dilaton, $\f_{0}
=-\ln\F_{0}$.

The relevant background quantities parameterized by $\o$ and $\l$ are
\bea 
H(t) & = & \frac{\sqrt{2\l}\, (\o +1)}{\sqrt{2\o +3}\,\sqrt{3\o +4}}\,\coth\,
(b\, t) \,\mp\, \frac{\sqrt{2\l}}{\sqrt{3} \,\sqrt{3\o +4}}\,\sinh^{-1}\, (b
t) \label{full1} \\ \nn \\
a(t) & = & a_{0}\, 2^{{\text \mp \frac{\sqrt{6\o +9}}{9\o +12}}}\, |\, b\, |^{
  -q(\o )}\,\Bigg|\sinh (b\, t)\Bigg|^{{\text \frac{\o +1}{3\o +4}}}
\,\left|\;\tanh\left(\,\frac{b\, t}{2}\,\right)\right|^{{\text \mp
    \frac{\sqrt{6\o +9}}{9\o +12}}} \label{full2} \\ \nn \\ 
\F_{0} (t) & = & \F^{\sscr (0)}_{0}\, 2^{{\text \pm \frac{\sqrt{6\o +9}}{3\o +4}}}\, |\, b\,
|^{ -r(\o )} \,\Bigg|\sinh (b\, t)\Bigg|^{{\text \frac{1}{3\o +4}}} \,
\left|\;\tanh\left(\,\frac{b\, t}{2}\,\right)\right|^{{\text \pm \frac{\sqrt{6\o
        +9}}{3\o +4}}} \label{full3} \ ,
\eea 
where we have used $q(\o )$ and $r(\o )$ as defined in~\refs{exponents1}.
Choosing $\o =-1$ these equations reduce to
\bea 
H(t) & = & \mp\sqrt{\frac{2\l}{3}}\,\sinh^{-1} \, (\sqrt{2\l } \, t)
\;\longrightarrow\; \mp\frac{1}{\sqrt{3}}\,\frac{1}{t}
\label{limit1} \\ \nn \\ 
a(t) & = & a_{0}\, 2^{{\text \mp\frac{1}{\sqrt{3}}}}\,
\Big|\,\sqrt{2\l}\,\Big|^{{\text
    \pm\frac{1}{\sqrt{3}}}}\,\left|\;\tanh\left(\frac{1}{2}\,\sqrt{2\l} \,
    t\right)\right|^{{\text \mp\frac{1}{\sqrt{3}}}}
\;\longrightarrow\; a_{0}\, \Big|\, t\,\Big|^{\;{\text \mp\frac{1}{\sqrt{3}}}}
\label{limit2} \\ 
\nn \\ \F_{0} (t) & = & \F^{\sscr (0)}_{0}\, 2^{{\scr \pm\sqrt{3}}}\,\Big|\,\sqrt{2\l}\,
\Big|^{{\scr -1 \mp\sqrt{3}}}\, \bigg|\sinh (\sqrt{2\l} \, t)\bigg| \,
\left|\;\tanh\left(\frac{1}{2}\,\sqrt{2\l} \, t\right)\right|^{{\scr
    \pm\sqrt{3}}} \label{limit3} \\ \nn \\ 
& \longrightarrow & \F^{\sscr (0)}_{0}\,\Big|\, t\,\Big|^{\;{\scr 1\pm\sqrt{3}}} \nn \ ,
\label{om=-1}
\eea
where the expressions after the arrow are valid in the limit $\sqrt{\l}\, t
\ll 1$. The prefactors in the equations~\refs{full2} and~\refs{full3} are
essentially the constants $\tilde{a}_{0}$ and $\tilde{\F}^{\sscr (0)}_{0}$ and have been
chosen such that in the special case $\o =-1, \l =0$ we get the simplest
possible initial values. All the upper signs correspond collectively to the
$(+)$ branch showing accelerated expansion for $t<0$ and the lower signs to
the $(-)$ branch with decelerated expansion for $t>0$. \\
One comment is in order concerning the sign of the potential. Up to now it was
assumed that $\l$ is positive; this is indeed not necessary: If $\l$ is
negative the correct expressions are obtained by substituting $b\rightarrow
|b |$ or $\l\rightarrow |\l |$ respectively and simultaneously replacing all
hyperbolic functions by their trigonometric counterparts.

\subsection{Mode functions for non zero cosmological constant}
\label{modeslinpot} 
Being equipped with the full expressions for the background with linearized
potential we are now in a position to investigate the influence of this
potential on the mode functions determined by equation~\refs{eom_modes}. It
possible to examine the dependence on the JBD parameter $\o$ equally well. It
is clear that the equation of motion for the mode functions in the
background~\refs{full1} -- \refs{full3} cannot be integrated analytically in
full generality. However, it is possible to examine the system in the limit
$\sqrt{\l}\, t \ll 1$. This can either be the small potential limit for
unconstrained time parameter or an expansion around $t=0$ valid for general
$\l$. In the small $\l$ limit it can be studied how the potential effectively
decouples from the solutions. Finally we will perform a numerical integration
of the most general equation~\refs{eom_modes} and compare the results with the
approximate formulae. We will later investigate the region of validity of the
stochastic approach. \\

First we want to give the asymptotic expressions of the solutions
to~\refs{eom_modes} in the limit $b\, t\gg 1$ for positive $\l$ (this is
equivalent to $\sqrt{\l}\, t\gg 1$ for all values of $\o$ we are interested
in); in this limit the functions $H(t)$ and $a(t)$ in~\refs{full1}
and~\refs{full2} can be simplified significantly. We find one exponentially
growing and one decaying mode, $\vf_{\bk}^{(1)}\sim e^{\,\O_{1}\, t}$ and
$\vf_{\bk}^{(2)}\sim e^{\,\O_{2}\, t}$ with the constants $\O_{1}$ and $\O_{2}$
given by
\be
\O_{1,2} = -\frac{1}{2}\,\left(\, 
\frac{3\,\sqrt{2\l}\, (\o +1)}{\sqrt{2\o +3}\sqrt{3\o +4}} \pm
\sqrt{8\l + \frac{18\l\, (\o +1)^{2}}{(2\o +3)(3\o +4)}}\;\right) \ .
\label{Omega}
\ee 
The two constants $\O_{1}$ and $\O_{2}$ are the {\em same} for {\em both}
branches. For $\o =-1$ this reduces to $\O_{1,2} = \pm\sqrt{2\l}$; for that
case we also find a correction,
\be
\O_{1,2} =\pm\sqrt{2\l\,\left(\, 1-\frac{|\bk |^{2}}{a_{0}^{2}}\, 2^{
      -1\pm{\text\frac{1}{\sqrt{3}}}}\,\l^{-1\mp{\text\frac{1}{\sqrt{3}}}}\,
  \right)} \ ,
\label{Omega2}
\ee
where the $\pm$ {\em under} the root distinguishes the $(+)$ branch (upper
signs) from the $(-)$ branch (lower signs) and the $\pm$ {\em in front} of the
root refers to the two different constants $\O_{1}$ and $\O_{2}$. We note that
similar asymptotic solutions can not be given for negative $\l$ because the
trigonometric functions are undetermined
at infinity. \\

Next we can try to solve equation~\refs{eom_modes} for small arguments $b\,
t\ll 1$ where the hyperbolic and trigonometric functions show the same
behaviour. The approximate equation of motion for the mode functions now read
\be
\ddot{\vf}_{\bk} \, + \, \frac{3q(\o)}{|\, t\, |}\,\dot{\vf}_{\bk} \, +
\,\left[ \frac{|\bk |^{2}}{a_{0}^{2}}\, |\, t\, |^{-2q(\o)} \, -
  \,\frac{2\l}{2\o +3} \right] \vf_{\bk} = 0 \ ,
\label{eom_modes-approx1}
\ee
and is valid for all $\o$ not too close to the singular point $-3/2$; an
overdot denotes a derivative with respect to $|\, t\, |$. The dependence on
$\l$ and its sign enters only through the last term in the brackets. It is
clear that the solutions only depend on the modulus of $t$. For $\bk =0$ we
obtain an expression which passes over to the zero mode in that limit and for
$\l =0$ we recover the free solution~\refs{modes1}.

Unfortunately this equation is still too difficult to solve in general the
trouble coming from the irrational powers of the first term in the bracket; it
cannot even be solved as a power law series. However, for the most interesting
case of $\o =-1$ on the $(+)$ branch the exponent can well be approximated by
$-2q^{(+)}\, (\o =-1) = 2/\sqrt{3}\approx 1.15$. So one can look for a
solution with the exponents $1$ and $2$ and compare the two cases. Later we
will comment on the validity of this simplification. We want to solve for $\o
=-1$ the equation
\be
\ddot{\vf}_{\bk} \, - \, \frac{\sqrt{3}}{|\, t\, |}\,\dot{\vf}_{\bk} \, +
\,\left[ \frac{|\bk |^{2}}{a_{0}^{2}}\, |\, t\, |^{m} \, -
  \,2\l \right] \vf_{\bk} = 0 \ ,
\label{eom_modes-approx2}
\ee
with $m$ replacing $1$ or $2$. This equation can be solved via the power law
series 
\be
\vf_{\bk} (t) = |\, t\, |^{r}\,\sum_{n=0}^{\infty} c_{n}\, |\, t\,|^{n}
\nn \ .
\label{powerlawansatz} 
\ee 
The exponent $r$ and the coefficients $c_{n}$ can be computed in the standard
way. For $m=1$ we get the two linearly independent solutions
\bea
\vf_{\bk}^{(1)} (t) & = & c_{0}\bigg[ 1
  - \frac{1+\sqrt{3}}{2}\,\l\,|\, t\,|^{2}
  - \frac{2+\sqrt{3}}{3}\frac{|\bk |^{2}}{a_{0}^{2}}\,|\, t\,|^{3} \nn \\
 & & \hspace{8mm} - \frac{3+2\sqrt{3}}{12}\,\l^{2}\,|\, t\,|^{4}
  - \frac{23+9\sqrt{3}}{390}\,\frac{|\bk |^{2}}{a_{0}^{2}}\,\l\, |\, t\,|^{5} 
  - \cdots\bigg] \label{powerlawsol1} \\
\vf_{\bk}^{(2)} (t) & = & c_{0}\;|\, t\,|^{1+\sqrt{3}}\;\bigg[ 1
  + \frac{3-\sqrt{3}}{6}\,\l\,|\, t\,|^{2}
  - \frac{4-\sqrt{3}}{39}\frac{|\bk |^{2}}{a_{0}^{2}}\,|\, t\,|^{3} \nn \\
 & & \hspace{24mm} + \frac{9-4\sqrt{3}}{132}\,\l^{2}\,|\, t\,|^{4}
  - \frac{381-157\sqrt{3}}{12870}\,\frac{|\bk |^{2}}{a_{0}^{2}}\,\l\,
          |\, t\,|^{5} 
  + \cdots\bigg] \ ,
\label{powerlawsol2}
\eea
and similarly for $m=2$, 
\bea
\vf_{\bk}^{(1)} (t) & = & \tilde{c}_{0}\bigg[ 1
  - \frac{1+\sqrt{3}}{2}\,\l\,|\, t\,|^{2}
  - \frac{3+2\sqrt{3}}{24}\,\Bigg( 2\l^{2} - (1-\sqrt{3} )\,\frac{|\bk |^{2}}
         {a_{0}^{2}}\,\Bigg)\, |\, t\,|^{4}  \nn \\
 & & \hspace{9mm}
  - \frac{21+13\sqrt{3}}{1584}\,\Bigg( 2\l^{2} - (7 - 3\sqrt{3} )\,
    \frac{|\bk |^{2}}{a_{0}^{2}}\,\Bigg)\,\l\, |\, t\,|^{6}
  - \cdots\bigg] \label{powerlawsol3} \\
\vf_{\bk}^{(2)} (t) & = & \tilde{c}_{0}\;|\, t\,|^{1+\sqrt{3}}\;\bigg[ 1
  + \frac{3-\sqrt{3}}{6}\,\l\,|\, t\,|^{2}
  + \frac{9-4\sqrt{3}}{264}\,\Bigg( 2\l^{2} - (3+\sqrt{3} )\,\frac{|\bk |^{2}}
         {a_{0}^{2}}\,\Bigg)\, |\, t\,|^{4}  \nn \\
 & & \hspace{24mm}
  + \frac{75-37\sqrt{3}}{36432}\,\Bigg( 2\l^{2} - (13 + 3\sqrt{3} )\,
    \frac{|\bk |^{2}}{a_{0}^{2}}\,\Bigg)\,\l\, |\, t\,|^{6}
  - \cdots\bigg] \ .
\label{powerlawsol4}
\eea
These are two linearly independent solutions for each case with two different
arbitrary constants $c_{0}$ and $\tilde{c}_{0}$ parameterising them.
Differences between the solutions for $m=1$ and $m=2$ are of third order in
$|\, t\, |$. It can be read off that the zero mode in the free case ($\l =0$)
is a constant or goes to zero as $\sim |\, t\,|^{1+\sqrt{3}}$ in accordance
with what we found in the previous discussion. Comparing with
equation~\refs{limit3} the two yet undetermined constants are fixed to be
$c_{0} = \tilde{c}_{0} = \F^{\sscr (0)}_{0}$. We are not interested in the
constant solution because we want to avoid a static universe.

Recursively, the higher order coefficients can be computed according to the
following relations:
\bea
c_{n} & = & \frac{2\l c_{n-2} - {\dis \frac{|\bk |^{2}}{a_{0}^{2}}}\, c_{n-3}}
                 {r^{2} + r\,(2n-1-\sqrt{3}) + n\, (n-1-\sqrt{3})}
\quad\mbox{for}\quad m=1 \nn \\ & & \\
c_{n} & = & \frac{2\l c_{n-2} - {\dis \frac{|\bk |^{2}}{a_{0}^{2}}}\, c_{n-4}}
                 {r^{2} + r\,(2n-1-\sqrt{3}) + n\, (n-1-\sqrt{3})}
\quad\mbox{for}\quad m=2 \nn \ . 
\label{recursive}
\eea


%
\begin{figure}[htb]
\begin{minipage}{7cm}
\centerline{(a)}
\psfig{figure=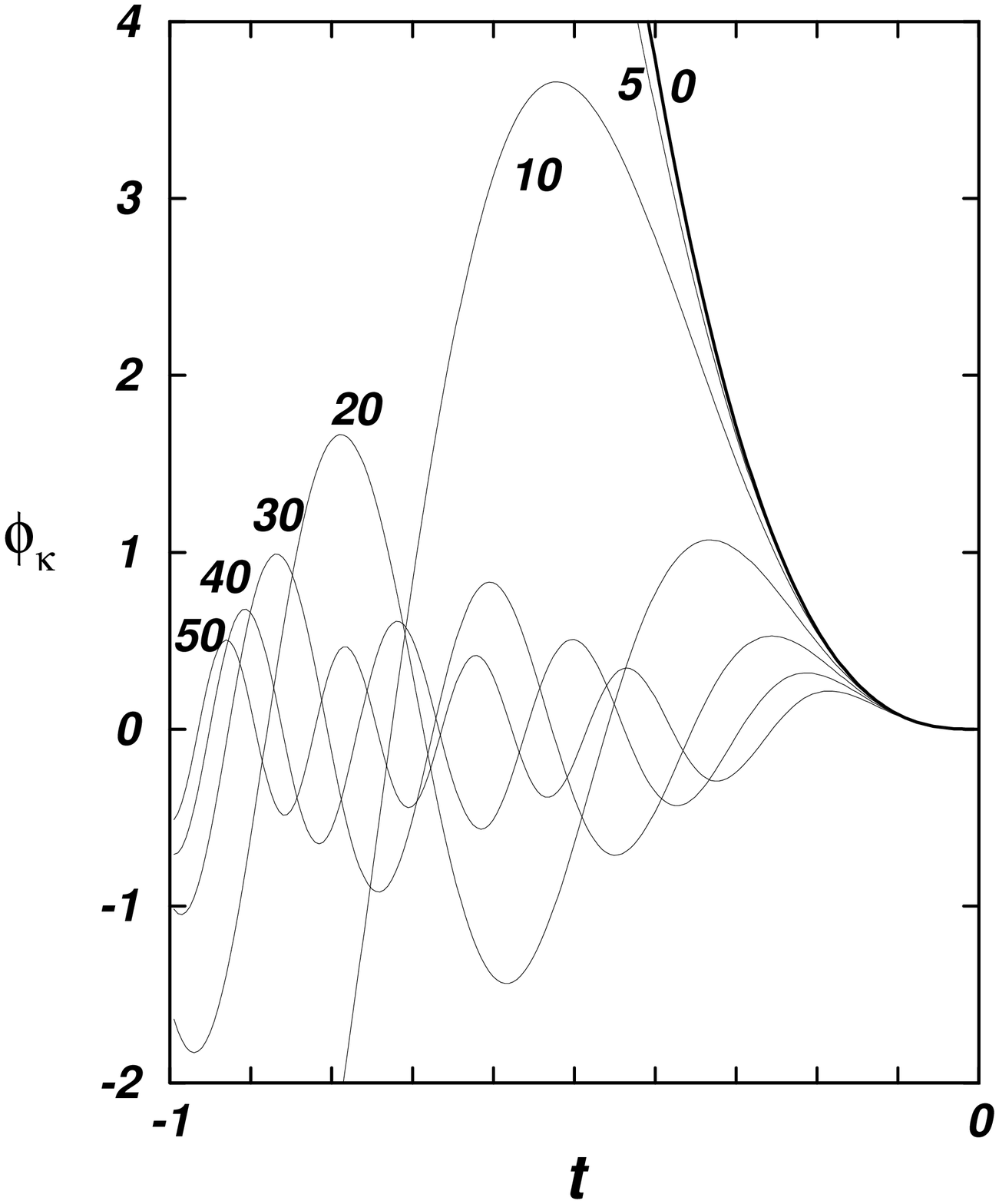,height=9cm,width=8cm,angle=0}
\end{minipage}\hfill
\begin{minipage}{7cm}
\centerline{(b)}
\psfig{figure=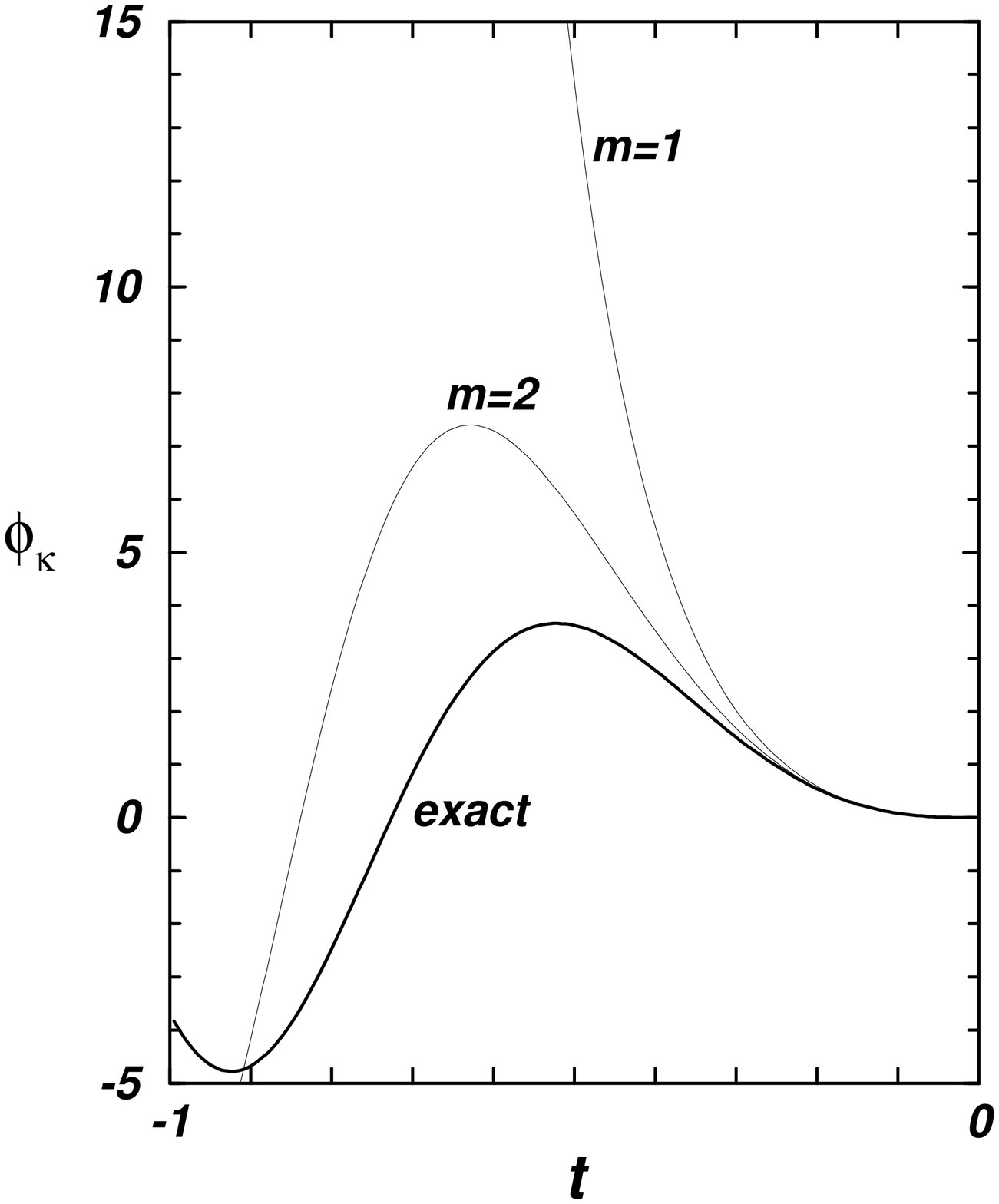,height=9cm,width=8cm,angle=0}
\end{minipage}
\caption{
  Mode functions for $\o =-1$. In (a) they are shown for various wavenumbers
  $|\bk |$ as indicated in the figure for negative times. The potential was
  set to $\l =1$. The zero mode corresponds to the bold line. In (b) the exact
  solution is compared with the approximate ones given in
  equation~\refs{powerlawsol2} and~\refs{powerlawsol4} for $\l =1$ and $|\bk
  |=10$. Accordance is achieved for small $t$. }
\label{fig1}
\end{figure}

%
\begin{figure}[htb]
\begin{minipage}{7cm}
\centerline{(a)}
\psfig{figure=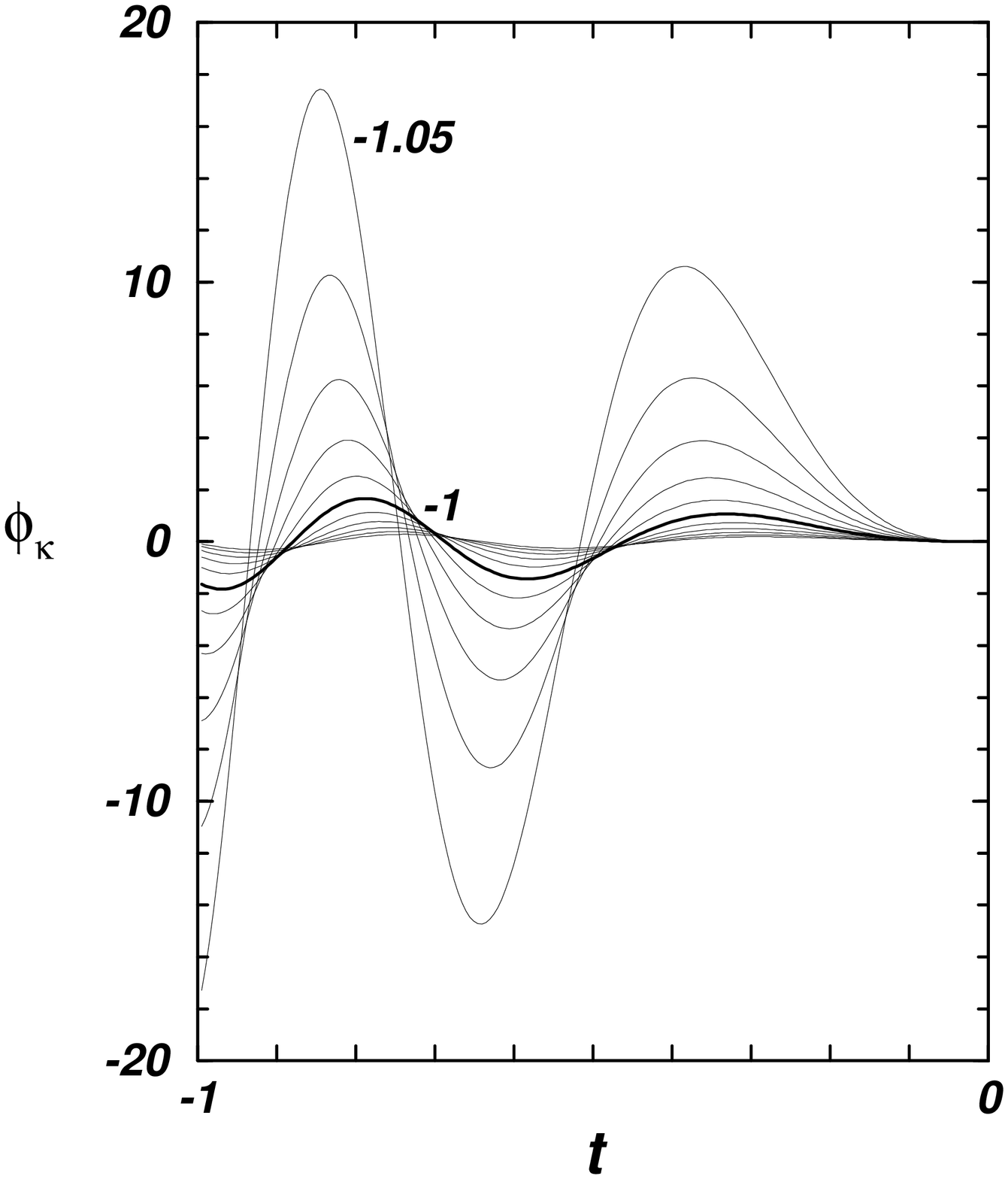,height=9cm,width=8cm,angle=0}
\end{minipage}\hfill
\begin{minipage}{7cm}
\centerline{(b)}
\psfig{figure=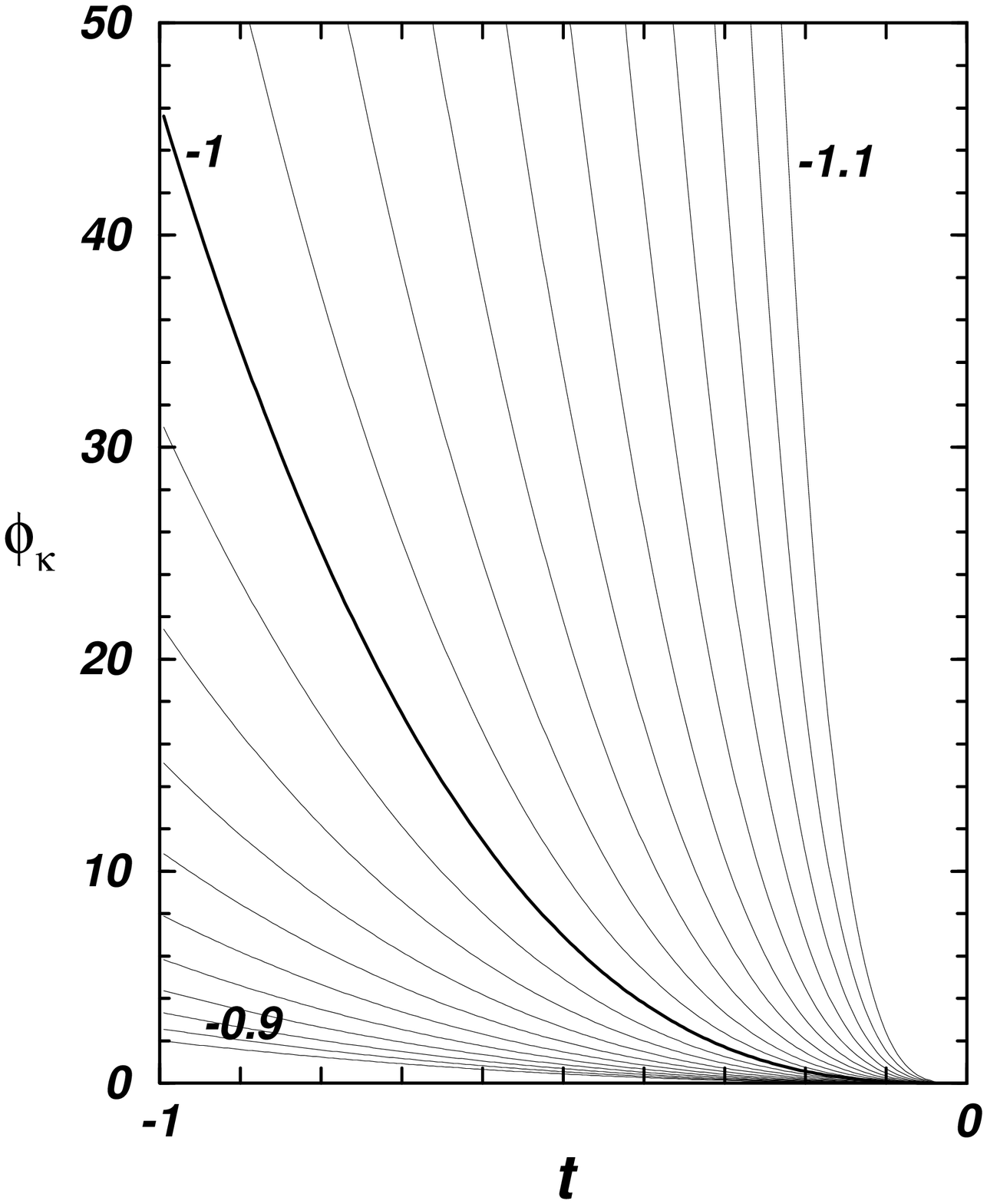,height=9cm,width=8cm,angle=0}
\end{minipage}
\caption{
  Mode functions for $\o \approx -1$. Their dependence on $\o$ is
  demonstrated. A sample of solutions for $\vf_{\bk}$ is depicted in (a) where
  $\l =1$ and $|\bk |=20$. The values of $\o$ vary between $-1.05$ and $-0.95$
  the bold line denoting $\o =-1$. In (b) $\l =1$ and $|\bk |=1$ has been
  chosen. }
\label{fig2}
\end{figure}
In figure~\ref{fig1} the results of a numerical integration of
equation~\refs{eom_modes} are shown for $\o =-1$ and the linear potential
discussed in this section. The qualitative picture is the same, however, for
general $\o$: Starting from a power law growth for the zero mode as $|\,
t\,|$ is getting larger, this behaviour is converted into oscillations with
increasing amplitude and frequency after a characteristic time for the mode
functions with larger $\bk$. This initial growth is well approximated by the
power law series solutions as can be seen in (b) where the full untruncated
solutions~\refs{powerlawsol2} and~\refs{powerlawsol4} are shown. The mode
functions are similar for all $\bk$ around the origin and $\vf_{\bk} /
\F_{0}\rightarrow 1$ as one approaches zero. We point out that for the {\em
  linear} potential it is not necessary that the higher mode functions are much
smaller than the zero mode, because in this case equation~\refs{eom_modes} is
exact and there are no corrections $\propto |\vf_{\bk}|^{2}$ etc.

Away from $\o =-1$ we shall be contented with the qualitative picture given by
the numerical integration shown in figure~\ref{fig2}. Although $\o$ is a free
parameter of the theory we are especially interested in $\o$ lying around
$-1$. Higher dimensional theories in $D$ dimensions compactified to a
maximally symmetric internal manifold yielding a $d+1$ dimensional
Kaluza--Klein cosmology can be viewed as a JBD theory described by a parameter
$\o =-1 + \frac{1}{D}$; the r\^ole of the JBD field is played by
$\sqrt{g_{\;mn}^{\sscr (D)}}$~\cite{freund}. Going from $\o =-1$ towards
somewhat larger values accessible to the above formula it is apparent that the
amplitude of the mode functions are significantly suppressed. They become
enormously large if $\o$ drops sightly below $-1$.


\subsection{Scalar field fluctuations}
\label{dispersion}
The presence of two different solutions, a generic feature of all JBD
theories, have their origin in the sign ambiguity in front of the square root
of equation~\refs{twobranches}. As long as only the classical theory is
considered the root is a well defined discriminant separating the two
branches. Fluctuations in the JBD field can, via backreactions, influence the
homogeneous degrees of freedom such that their behaviour has to be described
in terms of stochastic quantities as well (e.~g. mean value, variance). As
soon as the induced fluctuations in $H$ are comparable in size with the root
distinguishing the two classical paths it is conceivable that the system
undergoes a transition from one trajectory to the other and a strict 
distinction between the two trajectories present in the classical picture is 
no longer justified. \\ 

The subject of this section is to make this idea more precise. Let us begin
with a vanishing potential. According to our choice of the vacuum the mode
functions are given by $\vf_{\bk} = |\, t\, |^{-\a}\, H_{\n}^{(2)}(\b|\, t\,
|^{\g})$ together with equation~\refs{index} in which we shall always choose
the negative sign for $\n $ in order to deal with the cosmologically
interesting solutions only; in this case $\a + \g\n =-r$. The following
relations will turn out to be useful:
\bea
\frac{|\dot{\vf}_{\bk} |^{2}}{|\vf_{\bk} |^{2}} & = & \frac{r^{2}}{t^{2}} +
\frac{|\bk |^{2}}{a_{0}^{2}}\, t^{2\g -2}\, \left|\frac{H_{\n
      -1}^{(2)}\left(\b\, |\, t\, |^{\g}\,\right)}{H_{\n}^{(2)}\left(\b\, |\,
      t\, |^{\g}\,\right)}\right|^{2} +2r\,\frac{|\bk |}{a_{0}}\, |\, t\,
|^{\g -2}\, {\rm Re} \left( \frac{H_{\n -1}^{(2)}\left(\b\, |\, t\,
      |^{\g}\,\right)}{H_{\n}^{(2)}\left(\b\, |\, t\, |^{\g}\,\right)} \right)
\nn \\ 
\frac{\vf_{\bk} \dot{\vf}_{\bk}^{\ast} }{|\vf_{\bk} |^{2}} & = & \frac{r}{t}
+\frac{|\bk |}{a_{0}}\,\frac{1}{t}\, |\, t\, |^{\g}\, \frac{H_{\n
    -1}^{(1)}\left(\b\, |\, t\, |^{\g}\,\right)}{H_{\n}^{(1)}\left(\b\, |\,
    t\, |^{\g}\,\right)}
\label{pdotoverp} \\
\frac{\vf_{\bk}^{\ast} \dot{\vf}_{\bk}}{|\vf_{\bk} |^{2}} & = & \frac{r}{t}
+\frac{|\bk |}{a_{0}}\,\frac{1}{t}\, |\, t\, |^{\g}\, \frac{H_{\n
    -1}^{(2)}\left(\b\, |\, t\, |^{\g}\,\right)}{H_{\n}^{(2)}\left(\b\, |\,
    t\, |^{\g}\,\right)} \nn \ .
\eea 

As was demonstrated e.~g. in~\cite{yi} the variance in the fields
can be given, due to the stochastic differential equations, by the correlation
functions of the corresponding noise operators. They are given by
equation~\refs{correlator} where we shall adopt the specific choice $\ve=1$
\footnote{In an earlier analysis~\cite{yi} the semiclassical limit
  $\ve\rightarrow 0$ had to be chosen because it is only in this limit that
  the mode functions can be approximated by those of a massless free scalar
  field. Since we don't use such an approximation here we need not perform the
  limit $\ve\rightarrow 0$.}. For simplicity we neglect the spatial dependence
of the correlation functions (i.~e. we consider local fluctuations only) and
define the variance of the stochastic field operators to be the following 
positive definite expressions
\bea \langle\, \d\F^{2} \,\rangle & = & \left|\;\lim_{|\bf{x} -\bf{x}\pri
    |\rightarrow 0} \;\langle\, 0 \,|\,\chi\, ({\bf x}, t)\,\chi\, ({\bf x\pri
    }, t\pri )\, |\, 0 \,\rangle\;\;\right|_{t=t\pri \atop \ve =1} \nn \\ 
\langle\, \d v^{2} \,\rangle & = & \left|\;\lim_{|\bf{x} -\bf{x}\pri
    |\rightarrow 0} \;\langle\, 0 \,|\,\x\, ({\bf x}, t)\,\x\, ({\bf x\pri },
  t\pri )\, |\, 0 \,\rangle\;\;\right|_{t=t\pri \atop \ve =1} \label{disp1} \\ 
\langle\, \d\F\,\d v \,\rangle & = & \left|\;\lim_{|\bf{x} -\bf{x}\pri
    |\rightarrow 0} \;\langle\, 0 \,|\,\frac{1}{2}\,\left[ \chi\, ({\bf x},
    t)\,\x\, ({\bf x\pri }, t\pri ) + \x\, ({\bf x\pri }, t\pri )\,\chi\,
    ({\bf x}, t)\right] \,|\, 0 \,\rangle\;\;\right|_{t=t\pri \atop \ve =1}
\nn \\ \langle\, \d v\,\d\F\,\rangle & = & \left|\;\lim_{|\bf{x} -\bf{x}\pri
    |\rightarrow 0} \;\langle\, 0 \,|\,\frac{1}{2}\,\left[ \x\, ({\bf x},
    t)\,\chi\, ({\bf x\pri }, t\pri ) + \chi\, ({\bf x\pri }, t\pri )\,\x\,
    ({\bf x}, t)\right] \,|\, 0 \,\rangle\;\;\right|_{t=t\pri \atop \ve =1} \ 
, \nn \eea
where the cross terms have been defined in this way in order to avoid complex
quantities. Using equation~\refs{correlator} together with the given mode
functions we finally get with the help of~\refs{pdotoverp}
\bea
\langle\, \d\F^{2} \,\rangle & = & 
   \frac{1}{2\pi^{2}}\,\left|\frac{1-q}{q}\right|\, a^{3}\, H^{4}
   \left(\frac{\F_{0}}{\F_{0}^{(0)}}\right)\,
   \left|\, H_{\n}^{(2)} \left(\left|\text{\dis{\frac{q}{1-q}}}
   \right|\right)\right|^{2} \nn \\
\langle\, \d v^{2} \,\rangle & = &
   \langle\,\d\F^{2}\,\rangle\, H^{2}\left|\,\frac{r^{2}}{q^{2}}+ 
   \left|\,\frac{H_{\n -1}^{(2)}\left(\left|\dis{\frac{q}{1-q}}\right|\right)}
   {H_{\n}^{(2)}\left(\left|\dis{\frac{q}{1-q}}\right|\right)}\,\right|^{2} 
   \label{disp2} +\frac{2r}{|q|}\,{\rm Re}\left(\,
   \frac{H_{\n -1}^{(2)}\left(\left|\dis{\frac{q}{1-q}}\right|\right)}
   {H_{\n}^{(2)}\left(\left|\dis{\frac{q}{1-q}}\right|\right)}\,\right)\right| 
   \\
\langle\, \d\F\,\d v \,\rangle & = & 
   \langle\,\d\F^{2}\,\rangle\, |\, H\, |\,\left|\,\frac{r}{|q|}
   + {\rm Re}\left(\,
   \frac{H_{\n -1}^{(2)}\left(\left|\dis{\frac{q}{1-q}}\right|\right)}
   {H_{\n}^{(2)}\left(\left|\dis{\frac{q}{1-q}}\right|\right)}\,\right)\right|
   = \langle\, \d v\,\d\F\,\rangle \ . \nn
\eea
For the last equation the commutator~\refs{commutator} was used. The explicit
time dependence is reexpressed in terms of the background variables,
equation~\refs{V0solutions}. The above equations depend on an overall
parameter $a_{0}$ which we shall take to be of the order of unity. \\
An especially interesting example is the pre--big--bang scenario of
\cite{bru_ven} which is modeled by the choice $\o =-1$. In the following we
will consider the superinflationary $(+)$ branch realized by $q=-1/\sqrt{3} $
and negative $t$. As can be seen from the first line of the above system
together with the time dependence of the background quantities, the variance of
$\F$ grows for the $(+)$ branch like $(-t)^{-3}$ as one approaches the
curvature singularity at $t=0$. The additional factors of $H$ and $H^{2} $ in
the expressions for $\langle\, \d\F\,\d v \,\rangle $ and $\langle\, \d v^{2}
\,\rangle$ respectively make those variances grow even faster, i.~e.~
$\langle\, \d\F\,\d v \,\rangle\propto (-t)^{-4} $ and $\langle\, \d v^{2}
\,\rangle\propto (-t)^{-5} $ because the Hubble parameter scales like
$-q/(-t)$. Hence, the growth of the fluctuations is not bounded from above and
one can easily imagine them dominating over the classical evolution. In
particular, if $\d H_{rms} \equiv\langle\, \d H^{2} \,\rangle^{1/2}$, the
fluctuation of the Hubble parameter induced via backreactions, is of the same
order of magnitude as the square root of equation~\refs{twobranches} there is
no need to assume that the field will stay on its classical path. At this point
the field can be imagined to be part of either branch.

Next we will derive a quantitative estimate for the occurrence of this effect.
The quantity which distinguishes between the two branches is the expression
under the square root of equation~\refs{twobranches} which we will denote as $
A$,
\be
 A  = \frac{2\o +3}{12} \,\frac{\dot{\F}_{0}^{2}}{\F_{0}^{2}} 
- \frac{V(\F_{0})}{6\,\F_{0}} \ .
\label{alpha}
\ee
Note that $ A$ is completely given by the homogeneous field; we are looking for
its variation induced by substituting $\F_{0}\rightarrow\F_{0} +\d\F$ and
$v_{0}\rightarrow v_{0} +\d v$. The variation of $ A$ to second order gives
\bea
\d A & = & \frac{1}{6\F_{0}}\,\left(\frac{V(\F_{0})}{\F_{0}} - 
V\pri (\F_{0}) - (2\o +3)\frac{v_{0}^{2}}{\F_{0}^{2}}\right)\,\d\F 
\; + \;\frac{2\o +3}{6\F_{0}}\frac{v_{0}}{\F_{0}}\,\d v \nn \\
\label{varalpha}
 & + & \frac{1}{6\F_{0}}\,\Bigg( -\frac{V(\F_{0})}{\F_{0}^{2}}
+ \frac{V\pri (\F_{0})}{\F_{0}} - \frac{1}{2}\,V\dpr (\F_{0})
+ \frac{3(2\o +3)}{2}\,\frac{v_{0}^{2}}{\F_{0}^{3}}\,\Bigg)\,\d\F^{2} \\
 & + & \frac{(2\o +3)}{6\F_{0}}\,\frac{1}{2\F_{0}}\,\d v^{2}
\; - \;\frac{(2\o +3)}{6\F_{0}}\,\frac{2v_{0}}{\F_{0}^{2}}\,\d\F\d v \ . \nn
\eea
Since we neglected the spatial dependence of the correlation functions, the
above variation is due to local fluctuations of $\F$ and $v$ and it is
consistent to consider the expression~\refs{twobranches} with the spatial
derivatives dropped. The variance of the quantity $ A$ is
\bea
\langle\, \d A^{2} \,\rangle & = & \frac{1}{36\F_{0}^{2}}\,
\left(\frac{V(\F_{0})}{\F_{0}} - V\pri (\F_{0}) - 
(2\o +3)\frac{v_{0}^{2}}{\F_{0}^{2}}\right)^{2} \,
\langle\, \d\F^{2} \,\rangle
\; + \;\frac{(2\o +3)^{2}}{36\F_{0}^{2}}\,\frac{v_{0}^{2}}{\F_{0}^{2}}
\,\langle\, \d v^{2} \,\rangle \nn \\ 
 & + & \frac{(2\o +3)}{36\F_{0}^{2}}\,\frac{2v_{0}}{\F_{0}}\,
\left(\frac{V(\F_{0})}{\F_{0}} - V\pri (\F_{0}) - 
(2\o +3)\frac{v_{0}^{2}}{\F_{0}^{2}}\right)\,
\langle\, \d\F\d v \,\rangle \ .
\label{dispalpha}
\eea 
We shall identify $\langle\, \d\F^{2} \,\rangle$, $\langle\, \d v^{2}
\,\rangle$ and $\langle\, \d\F\d v \,\rangle$ with the expressions given in
equation~\refs{disp2}. As a condition for an overlap of the two distinct
classical trajectories induced by stochastic effects we require $\d A_{rms} / A
\sim 1$. For general $\o$ and a vanishing potential we get
\be
\frac{\d A_{rms}}{ A} =
2\,\sqrt{\frac{\langle\, \d\F^{2} \,\rangle}
{\F_{0}^{2}} + \frac{\langle\, \d v^{2} \,\rangle}{v_{0}^{2}} -
2\frac{\langle\, \d\F \,\d v \,\rangle}{\F_{0} v_{0}} } \ .
\label{flucalpha}
\ee
In the model considered so far $\d A_{rms} / A \simeq \d H_{rms} / H$ (apart
from a trivial factor $2$ coming from the square root) which means that right
at the point when an overlap becomes significant a large fraction of the total
energy is stored in fluctuations and spatial gradients may no longer be
neglected. Although no assumption is necessary concerning the ratio $\vf_{\bk}
/ \F_{0}$ for a vanishing or a linearized version of the potential the energy
density due to the fluctuations should not be dominating over the
contribution coming from the background. From this point of view the condition
$\d A_{rms} / A \sim 1$ will be met at the border of validity of this method.

Setting $\o =-1$ we find a quantitative estimate by inserting the expressions 
derived in equation~\refs{disp2} into~\refs{flucalpha},
\bea
\frac{\d A_{rms}}{ A} & = &
\frac{1}{\pi}\,\frac{\sqrt{2}}{\sqrt{3+\sqrt{3}}}\,\left(\,\F_{0}^{(0)}\,\right)^{\;-\frac{1}{2}}
\left(H^{2}\, a^{\frac{3}{2}} \,\F_{0}^{-\frac{1}{2}}\right) \, \left|\, H_{\n
    -1}^{(2)} \left(\dis{\frac{1}{1+\sqrt{3}}} \right)\right| \nn \\ 
& = & \frac{1}{3\pi}\,\frac{\sqrt{2}}{\sqrt{3+\sqrt{3}}}\,
\frac{a_{0}^{\frac{3}{2}}}{\F_{0}^{(0)}}\, \left|\, H_{\n -1}^{(2)}
  \left(\dis{\frac{1}{1+\sqrt{3}}} \right)\right|\,
(-t)^{-\left(\text{\frac{5}{2}}\, +\sqrt{3}\right)}
   \label{dalpharms} \ , 
\eea
with the index of the Hankel function given by
$\n=-\frac{1}{2}\frac{3+\sqrt{3}}{1+\sqrt{3}}$ and $\left|\, H_{\n -1}^{(2)}
  \left(\text{\frac{1}{1+\sqrt{3}}} \right)\right|\approx 7.49$. In the second
line we have used the time dependence of the background fields for the
pre--big--bang branch. The result depends on the initial values $a_{0}$ and
$\F_{0}^{(0)}$ which we assume to be of the order of one. If we had not made
the choice $\ve =1$ at the beginning the result would also depend on
$\ve^{3/2}$ which can always be absorbed by the initial values. No matter what
the prefactors are, the strong time dependence with the power of
$-\frac{5}{2}\, -\,\sqrt{3}\approx -4.23$ ensures that the condition $\d
A_{rms} / A \sim 1$ is unavoidably met at a time which can be read off from
equation~\refs{dalpharms}; this time is $\sim 1$ if the initial
condition $a_{0}^{3/2}\, /\,\F_{0}^{(0)}$ is of the order of one. \\

The same as was demonstrated above for the stringy example $\o =-1$ can very
easily be generalized for arbitrary $\o$ (again restricting ourselves to a
vanishing potential). Equation~\refs{flucalpha} is valid for general $\o$. The
structure of $\d A_{rms} / A $ is very similar to that of~\refs{dalpharms},
\be
\frac{\d A_{rms}}{ A} = 
\frac{\sqrt{2}}{\pi}\, \sqrt{\left|\frac{1-q(\o)}{q(\o)}\right|}\, \left|\,
  H_{\n }^{(2)} \left(\,\left|\frac{q(\o)}{1-q(\o)}
    \right|\,\right)\right|\, F(\o )\,
\left(\,\F_{0}^{(0)}\,\right)^{\;-\frac{1}{2}} \left(H^{2}\, a^{\frac{3}{2}}
  \,\F_{0}^{-\frac{1}{2}}\right)
\label{dalpharms2} \ , 
\ee with $F(\o )$ denoting a function of $\o$ which is approximately one in
the region of interest. Using the general background solutions for zero
potential~\refs{V0solutions} and~\refs{exponents1} it is obvious that $\d
A_{rms} / A \sim (-t)^{-2 + \frac{3}{2}q(\o) - \frac{1}{2}r(\o)}$ for the
$(+)$ branch. The exponent, being a function of $\o$, is always negative (as
it should be) and a monotonically growing function increasing from
$-\frac{5}{2}\, -\,\sqrt{3}$ at $\o =-1$ to $-\frac{5}{2}$ at $\o =0$; this is
again a sufficiently fast growing function for the value of $\d A_{rms} / A $
becoming one at a time which is of the order of unity. The $\o$~--~dependent
prefactor of equation~\refs{dalpharms2} shows the remarkable feature that it
has a singular point at $\o =0$ where it becomes plus infinity. This means
that approaching $\o =0$ from either direction drastically enhances the ratio
$\d A_{rms} / A $ such that the overlap condition is easily fulfilled. \\ 
In trying to explore the influence of a linearized potential on the ratio $\d
A_{rms} / A $ one would have to refer to the approximate mode functions we
found in~\refs{powerlawsol2} and~\refs{powerlawsol4} for $\o =-1$ and insert
them into the corresponding equation which is similar to~\refs{flucalpha}.
>From the resulting formulae it is, however, hard to see how a small but non
zero potential influences the effect described above. Therefore, we shall
rather directly integrate the coupled stochastic equations of
motion~\refs{stoeom1}, \refs{stoeom2} and~\refs{stochHubble} using the
approximate mode functions valid for $\sqrt{\l}\, t < 1 $. The discretized
version of the equations can be written in the form
\bea
\Phi_{n+1} & = & \Phi_{n} + v_{n}\,\d t +
\frac{1}{3\pi}\,\sqrt{\frac{1+\sqrt{3}}{2}}\, |\F^{\sscr (0)}_{0} |\,
a_{0}^{\text{\frac{3}{2}}} \,\left(\, 1+\frac{3-\sqrt{3}}{6}\,\l \,|\, t_{n}\,
  |^{2}\,\right)\,\times
\nn \\
& & |\, t_{n}\, |^{-1+\text{\frac{\sqrt{3}}{2}}}\,\sqrt{\d t}\; w_{n}^{(1)} 
\label{discrete1} \\
\nn \\ 
v_{n+1} & = & v_{n} -3\, H_{n} v_{n}\,\d t + 2\l\,\Phi_{n}\,\d t +
\frac{1}{3\pi}\,\frac{(1+\sqrt{3})^{\text{\frac{3}{2}}}}{2}\, 
|\F^{\sscr (0)}_{0} |\,
a_{0}^{\text{\frac{3}{2}}}\,\times \nn \\ & & \left(\,
  1+\left(\,\frac{3-\sqrt{3}}{6} +
    \frac{1}{3}\,\frac{3-\sqrt{3}}{1+\sqrt{3}}\,\right)\,\l \,|\, t_{n}\,
  |^{2}\,\right)\, |\, t_{n}\, |^{-2+\text{\frac{\sqrt{3}}{2}}}\,\sqrt{\d t}\;
w_{n}^{(2)} \label{discrete2} \\ \nn \\
H_{n} & = & -\frac{1}{2}\frac{v_{n}}{\Phi_{n}} \pm 
\sqrt{\frac{1}{12}\frac{v_{n}^{2}}{\Phi_{n}^{2}} + \frac{\l}{3}} \ ,
\label{discrete3}
\eea
with $n$ being the step number and $\d t$ the step size. To simulate a
discrete Wiener process we use the two independent random Gaussian deviates
$w^{(1)}_{n}$ and $w^{(2)}_{n}$ of unit variance and zero mean,
\be 
\langle\, w^{(i)}_{n}\,\rangle =0 \qquad\mbox{and}\qquad 
\langle\, w^{(i)}_{m}\,w^{(j)}_{n}\,\rangle = \d_{ij}\,\d_{mn} \ .
\label{var}
\ee 
The above discretized equations are in agreement with It\^o's
definition~\cite{ito} of the stochastic noise. It is assumed that other
interpretations of the noise term have only small effects on the qualitative
behaviour of the system~\cite{yi}. The Gaussian deviates are produced by a
standard method~\cite{recipes} from a very well tested random number
generator. \\

\begin{figure}[htb]
\begin{minipage}{7cm}
\centerline{(a)}
\psfig{figure=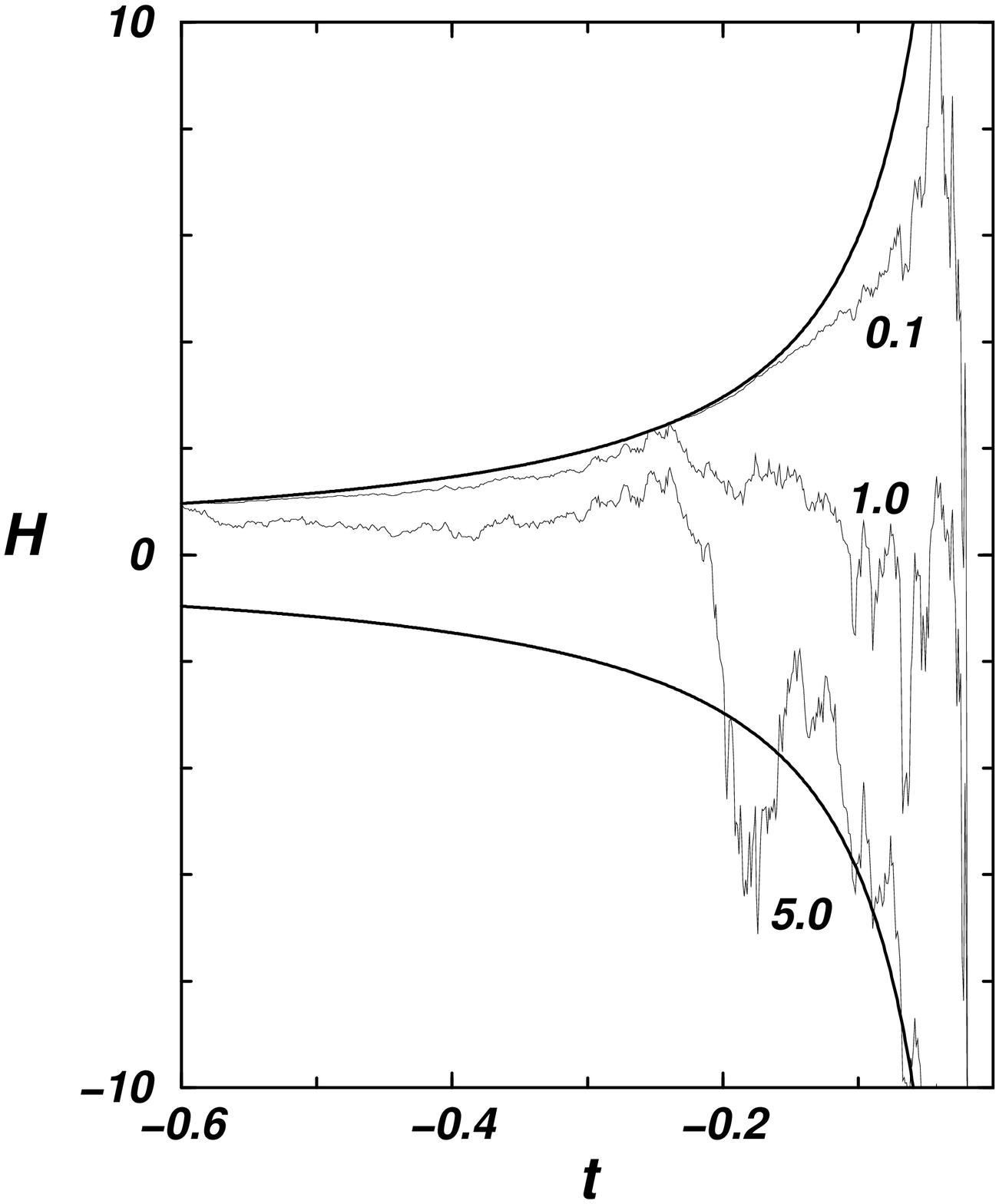,height=9cm,width=8cm,angle=0}
\end{minipage}\hfill
\begin{minipage}{7cm}
\centerline{(b)}
\psfig{figure=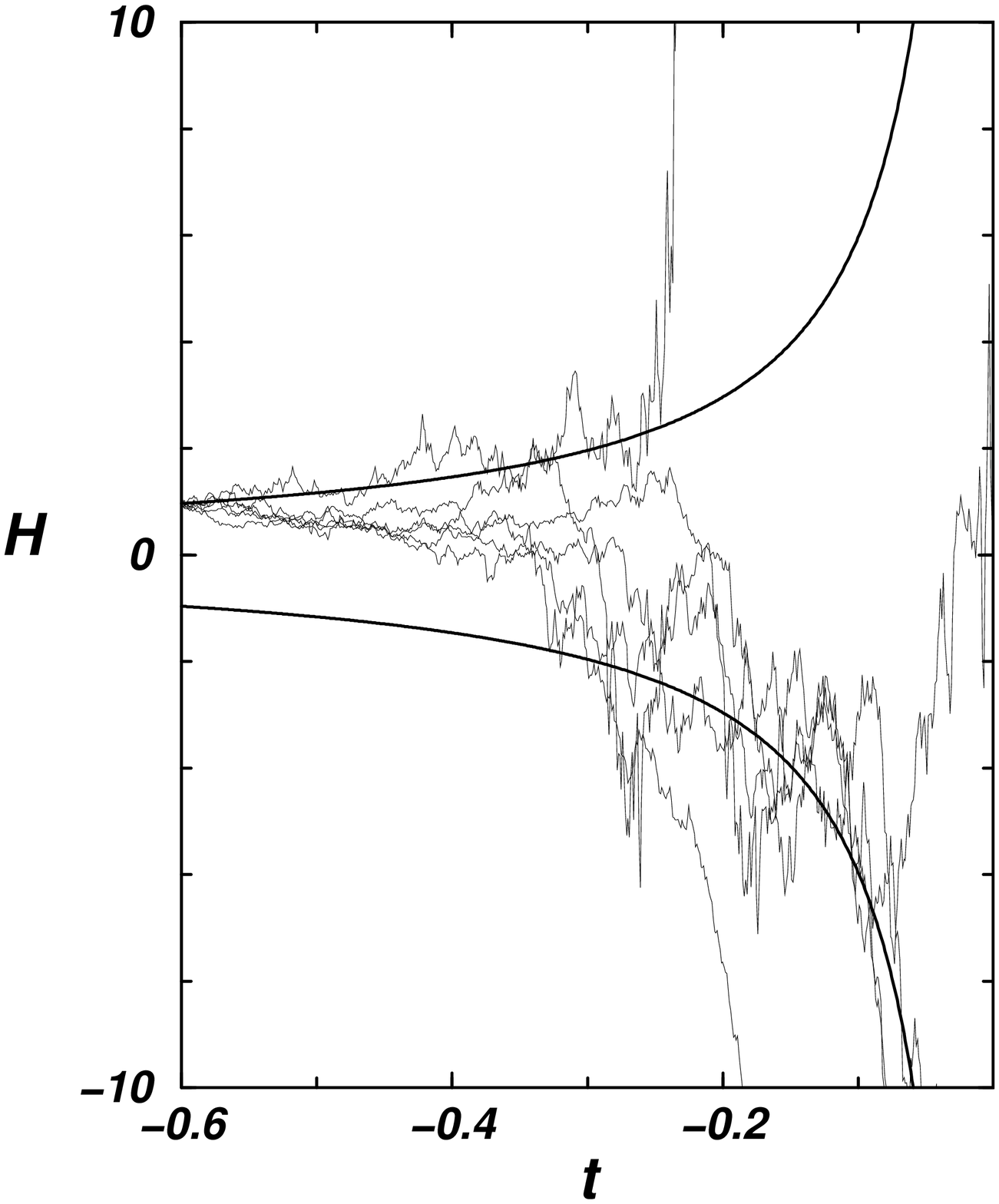,height=9cm,width=8cm,angle=0}
\end{minipage}
\caption{The numerical integration of the stochastic equations of 
  motion~\refs{discrete1} -- \refs{discrete3} for the $(+)$ branch of the $\o
  =-1$ theory are shown ($\l =1$ has been chosen). In both plots the result of
  the classical equations (bold lines) are compared with the stochastic
  trajectories. The Hubble parameter is given as a function of (negative) time.
  (a) Various stochastic paths parameterized by the effective fluctuation
  strength $|\F^{\sscr (0)}_{0} |\, a_{0}^{3/2}$ (as indicated) are compared.
  (b) The specific path "$5.0$" of (a) for different randomly chosen
  initializations of the random number generator. All stochastic trajectories
  emerge from the same starting point being also the starting point of the
  upper ($H>0$) classical solution.}
\label{fig3}
\end{figure}

As an example, the results of the numerical integration of the
equations~\refs{discrete1} -- \refs{discrete3} are shown in figure~\ref{fig3}.
The integration was done for the $(+)$ branch (referring to the lower sign in
equation~\refs{discrete3}) and negative $t$. Both plots show two classical
trajectories (obtained by switching off the stochastic noise terms)
corresponding to accelerated expansion ($H>0$) and accelerated contraction
($H<0$). On the classical level there is no transition possible between the
regions of positive and negative $H$. In both examples the initial conditions
of all stochastic trajectories are chosen to emerge from the same starting
point which is also the starting point of the classical path for $H>0$. In (a)
the free parameter $|\F^{\sscr (0)}_{0} |\, a_{0}^{3/2}$ is varied which
effectively sets the strength of the fluctuations. While smoothly turning on
the fluctuations by increasing the value of $|\F^{\sscr (0)}_{0} |\,
a_{0}^{3/2}$ the stochastic trajectory bends towards the second classical path
characterized by negative $H$. Phenomenologically this situation is of course
not attractive; on the other hand this provides an example that via
backreactions two a priori disconnected regions in the space of solutions to
the classical equations of motion are no longer separated when sizable
fluctuations enter the analysis.
\\
Certainly not all stochastic trajectories behave as the one depicted in (a).
In plot (b) the path labelled by "$5.0$" in (a) is redrawn together with other
realizations, i.~e. all the parameters are held fixed but the initialization of
the random number generator is different. The dispersion of the whole ensemble
of stochastic trajectories grows much faster than the difference between the
two distinct classical solutions; this is in accordance with what we found
analytically in this subsection.


%
The condition $\sqrt{\l}\, t\ll 1$ is fulfilled at any instance of the
integration; moreover, the relative size of the fluctuations are always kept
under control during the integration meaning that $\d\Phi / \Phi\ll 1$ and $\d
v / v \ll 1$ are satisfied for almost all points; thus violently fluctuating
intervals where the whole procedure becomes doubtful are excluded. As was
already anticipated in~\cite{bru_ven} on the classical level, we again find
from the numerical integration that the influence of a linear potential on the
dynamics during the phase of kinetic inflation is negligible also when
fluctuations are included. The drastic deviations of the stochastic
trajectories from the classical ones are present also for zero potential and
are qualitatively similar. From equations~\refs{discrete1}
and~\refs{discrete2} it follows that the fluctuations in $v$ dominate over
those in $\Phi$ by one power in $t$ as one approaches $t=0$. Indeed we find
from the numerics that the stochastic noise term in equation~\refs{discrete1}
is subdominant. \\

On the basis of the intuitions associated with the standard inflationary 
paradigm one would expect  the size of the fluctuations to  be set by the Hubble
parameter alone. However, in our case we can see that the results 
depend somewhat on the
parameter $|\F^{\sscr (0)}_{0} |\, a_{0}^{3/2}$. 
This is not too surprising, as there is another mass scale in the problem,
namely the Planck scale which is actually the dynamical variable equivalent to 
$\F$, hence the dependence on initial conditions can be seen as the dependence on the 
initial value of the effective Planck scale. Since there is no
canonical way of determining a unique, internal set of initial conditions for
the pre--big--bang epoch, the size of the fluctuations can only be given in
terms of these parameters. The qualitative picture remains however unchanged no
matter what their specific values might be, as long as they are not too exotic.
From figure~\ref{fig3} one can draw the conclusion that the choice of the
initial conditions merely fixes the instant of time, when the stochastic
trajectory bends away from the original classical solution.

Finally,  we should comment on the assumed value of  the 
splitting parameter $\ve =1$.
In investigations of potential driven inflationary scenarios with 
a period very close to the actual de Sitter epoch one tends to go 
with the value of  $\ve $ to zero. This is usually very helpful in simplifying 
calculations, but in our opinion not so well justified on physical grounds, as 
strictly speaking in this limit all the Fourier modes except the zero mode 
are considered ``fluctuations''. Hence here we decided to keep  $\ve $ finite 
and of the order one which singles out the causal horizon as the splitting 
scale at any instant of the evolution. 
It is clear that $\ve $ could
have been chosen smaller - this would mean that we shift more and more degrees of 
freedom into the fluctuating part of the field. It turns out that over a wide range of
 finite  $\ve $ the results of the stochastic analysis are indeed physically 
equivalent.

\section{Summary and Conclusion}
\label{conclusion}
Jordan--Brans--Dicke (JBD) theories with a linearized potential for the JBD
field are investigated in the framework of a stochastic analysis which is
capable of taking fluctuations of this field into account; their backreactions
on the classical background are examined. We split the JBD scalar in sub-- and
super--horizon parts treating the long wavelength modes as the background
quantity whose time evolution is subject to the short wavelength fluctuations.
We derive the stochastic equations of motion for the system together with the
two point correlation functions of the stochastic noise operators; they finally
set the strength of the random force term in the Langevin equations
which, in principle, can be integrated for any potential. \\
For a non vanishing potential we give the scalar field mode functions for
several limiting cases; they finally enter the expressions for the correlation
functions. We compute the mode functions in the limit of $\sqrt{\l}\, t\gg 1$
for general $\o$ and give the approximate solutions for $\sqrt{\l}\, t\ll 1$
for the $\o =-1$ model in terms of power law series; this allows one to study
how the mode functions are modified if a small but finite potential is switched
on. The mode functions can not be given analytically in the most general case;
Therefore, we integrate the system numerically and analyze the influence of
general $\l$ and $\o$. We find that all mode functions are regular at $t=0$ and
turn from a power law growth to an oscillatory regime after a time $t$ given by
the ratio $|\bk |/\l$. Moreover, we demonstrate that they strongly depend on
the value of $\o$ chosen with a significant increase
when  $\o$ drops even slightly below $-1$. \\
After defining the variances of the random variables in terms of the
correlation functions we explicitly give their form  in the zero potential limit and
for the stringy model $\o =-1$. Arguing that two distinct classical solutions can
no longer be distinguished as soon as the dispersion of the fluctuations is of
the same order of magnitude as the separation between the two classical
trajectories, we show for general $\o$ that this condition is met for all cases
as $t=0$ is approached. We point out that this fact is not due to the mode
functions (since they remain finite for $t\rightarrow 0$), but due to a
combination of the background fields which gets singular close to $t=0$. 
We find a strong enhancement of this effect for $\o\rightarrow
0$. This is where many theories of gravity arising from Kaluza--Klein theories
after compactification to four dimensions are located~\cite{freund}.

Finally, for $\o =-1$, we perform an integration of the stochastic equations of
motion for a finite potential demonstrating how the stochastic ensemble
corresponding to the full quantum scalar field evolves under the influence of
the scalar field fluctuations. The dynamics is such that the ensembles
representing classical solutions which belong to disconnected solutions start
overlapping. We again find that the dispersion of the fluctuations grows to
achieve the magnitude of the term separating the two classical solutions. This
phenomenon can be interpreted as the quantum mechanical realization of
connecting classically disconnected solutions at the level of field theoretical
considerations.

\section*{Acknowledgment}
This work was supported by the "Sonderforschungsbereich 375-95 f\"ur
Astro-Teilchenphysik der Deutschen Forschungsgemeinschaft", European
Commission TMR programs \\ ERBFMRX-CT96-0045 and ERBFMRX-CT96-0090, and 
by Polish Commitee for Scientific Research grant 2 P03B 040 12. 

It is a
pleasure to thank S.~Theisen and A.~Lukas for many interesting discussions.


%

\newpage
\begin{thebibliography}{99}
\bibitem{brans} P. Jordan, {\em Z. Phys.} {\bf 157} (1959) 112; C. Brans and
  R.~H. Dicke, {\em Phys. Rev.} {\bf 124} (1961) 925.

\bibitem{reasenberg} R.~D. Reasenberg {\em et al.}, {\em Astrophys. J.} {\bf
  234} (1979) L219.

\bibitem{levin} J.~J. Levin, {\em Phys. Rev. D} {\bf 51} (1995) 1536.

\bibitem{freund} P.~G.~O. Freund, {\em Nucl. Phys. B} {\bf 209} (1982) 146. 

\bibitem{eff_action} E.~S. Fradkin and A.~A. Tseytlin, {\em Phys. Lett. B}
  {\bf 158} (1985) 316; {\em Nucl. Phys. B} {\bf 261} (1985) 1; C. Callan, D.
  Friedan, E. Martinec and M. Perry, {\em Nucl. Phys. B} {\bf 262} (1985) 593.

\bibitem{inflation} M.~S. Berman and M.~M. Som, {\em Phys. Lett. A} {\bf 136}
  (1989) 206; M.~S. Berman, {\em Phys. Lett. A} {\bf 142} (1989) 335; A.
  Linde, {\em Phys. Lett. B} {\bf 249} (1990) 18; J. McDonald, {\em Phys.
    Rev. D} {\bf 44} (1991) 2314.

\bibitem{la} D. La and P.~J. Steinhardt, {\em Phys. Rev. Lett.} {\bf 62}
  (1989) 376.

\bibitem{ven} G. Veneziano, {\em Phys. Lett. B} {\bf 265} (1991) 287; M.
  Gasperini and G. Veneziano, {\em Astropart. Phys.} {\bf 1} (1993) 317.

\bibitem{bru_ven} R. Brustein and G. Veneziano, {\em Phys.
    Lett. B} {\bf 329} (1994) 429. 

\bibitem{kal_mad} N. Kaloper, R. Madden and K.~A. Olive, {\em Nucl. Phys. B}
  {\bf 452} (1995) 677.

\bibitem{east_maed} R. Easther and K. Maeda, {\em Phys. Rev. D} {\bf 54}
  (1996) 7252.

\bibitem{quantcos} see e.~g. M.~C. Bento and O. Bertolami, {\em Class. Quant.
    Grav.} {\bf 12} (1995) 1919. M. Gasperini, J. Maharana and G. Veneziano,
  {\em Nucl. Phys. B} {\bf 472} (1996) 349. M. Gasperini and G. Veneziano,
  {\em Gen. Relativ. Gravit.} {\bf 28} (1996) 1301; see also M.~P.
  Dabrowski and C. Kiefer, {\em Phys. Lett. B} {\bf 397} (1997) 185.

\bibitem{odd} K.~A. Meissner and G. Veneziano, {\em
    Phys. Lett. B} {\bf 267} (1991) 33; A. Sen, {\em Phys. Lett. B} {\bf 279}
  (1991) 295; M. Gasperini and G. Veneziano, {\em Phys. Lett. B} {\bf 277}
  (1992) 256; J. Maharana, {\em Phys. Lett. B} {\bf 296} (1992) 65; C. Klimcik
  and A.~A. Tseytlin, {\em Phys. Lett. B} {\bf 323} (1994) 305.

\bibitem{lidsey} J.~E. Lidsey, {\em Phys. Rev. D} {\bf 55} (1997)
  3303.

\bibitem{garcia} J. Garcia--Bellido, {\em Nucl. Phys. B} {\bf 423} (1994) 221;
  J. Garcia--Bellido, A. Linde and D. Linde, {\em Phys. Rev. D} {\bf 50}
  (1994) 730; A.~A. Starobinsky and J. Yokoyama, {\em Proceedings of the
    $4^{th}$ Workshhop on General Relativity and Gravitation}, Kyoto, Japan,
  28 Nov -- 1 Dec 1994; M. Susperregi, {\em Phys. Rev. D} {\bf 55} (1997) 560.

\bibitem{mallik} S. Mallik and D. Rai Chaudhuri, {\tt hep-ph/9612427}.

\bibitem{gravrad} R. Brustein, M. Gasperini, M. Giovannini, V.~F. Mukhanov and
  G. Veneziano, {\em Phys. Rev. D} {\bf 51} (1995) 6744; M. Gasperini and M.
  Giovannini, {\em Phys. Lett. B} {\bf 282} (1992) 36; M. Gasperini and M.
  Giovannini, {\em Phys. Rev. D} {\bf 47} (1993) 1519; J. Hwang, {\tt
    hep-th/9608041}.

\bibitem{antoniadis} I. Antoniadis, J. Rizos and K. Tamvakis, {\em Nucl. Phys.
    B} {\bf 415} (1994) 497; E. Kiritsis and C. Kounnas, {\em Phys. Lett. B}
  {\bf 331} (1994) 51; S.--J. Rey, {\em Phys. Rev. Lett.} {\bf 77} (1996)
  1929.

\bibitem{rey} S.--J. Rey, {\em Nucl. Phys. B} {\bf 284} (1987) 706, and
  references therein.

\bibitem{risken} H. Risken: {\em The Fokker--Planck Equation,} Second Edition,
  Springer--Verlag, Berlin, 1989; N. G. van Kampen: {\em Stochastic Processes
    in Physics and Chemistry,} North--Holland Publishing Company, Amsterdam,
  1981.

\bibitem{yi} I. Yi and E.~T. Vishniac, {\em Phys. Rev. D} {\bf 47} (1993)
  5280, and references therein.

\bibitem{uehara} K. Uehara and C.~W. Kim, {\em Phys. Rev. D} {\bf 26} (1982)
  2575. 

\bibitem{ito} K. It\^o, {\em Proc. Imp. Acad.} {\bf 20} (1944) 519.

\bibitem{recipes} W. H. Press, B. P. Flannery, S. A. Teukolsky and W. T.
  Vetterling: {\em Numerical Recipes,} Cambridge University Press, New York,
  1987.

\end{thebibliography}
\end{document}